\documentclass[prd,showkeys,floatfix,twocolumn,amsmath,amssymb,floatfix]{revtex4}
\usepackage{graphicx,color,dcolumn,booktabs,bm}
\usepackage{longtable,lscape}
\usepackage{amssymb}
\usepackage{amsmath}
\usepackage{indentfirst}
\usepackage{epsfig}
\usepackage{feynmf}
\usepackage{epstopdf}
\usepackage{subfigure}
\usepackage{slashed}
\usepackage{cases}
\usepackage{multirow}
\usepackage[colorlinks,citecolor=blue,anchorcolor=red,menucolor=red,linkcolor=red,filecolor=red,runcolor=red,urlcolor=blue,frenchlinks=red]{hyperref}

\allowdisplaybreaks[3]
\begin{document}
\title{Light tetraquark states with $J^{PC}=1^{--}$ from QCD sum rules}
%
\author{Yi-Wei Jiang$^1$}
\email{yiweijiang@seu.edu.cn}
\author{Hua-Xing Chen$^1$}
\email{hxchen@seu.edu.cn}
\author{Er-Liang Cui$^2$}
\email{erliang.cui@nwafu.edu.cn}
\author{Ding-Kun Lian$^1$}
\email{liandk@seu.edu.cn}
\author{Wen-Ying Liu$^3\,^4\,^5\,^6$}
\email{liuwenying@lzu.edu.cn}
\author{Niu Su$^7$}
\email{suniu@tyut.edu.cn}

\affiliation{
$^1$School of Physics, Southeast University, Nanjing 210094, China
\\
$^2$College of Science, Northwest A\&F University, Yangling 712100, China
\\
$^3$School of Physical Science and Technology, Lanzhou University, Lanzhou 730000, China
\\
$^4$Lanzhou Center for Theoretical Physics, Key Laboratory of Theoretical Physics of Gansu Province, Key Laboratory of Quantum Theory and Applications of MoE, Gansu Provincial Research Center for Basic Disciplines of Quantum Physics, Lanzhou University,
Lanzhou 730000, China
\\
$^5$MoE Frontiers Science Center for Rare Isotopes, Lanzhou University, Lanzhou 730000, China
\\
$^6$Research Center for Hadron and CSR Physics, Lanzhou University and Institute of Modern Physics of CAS, Lanzhou 730000, China
\\
$^7$College of Physics and Optoelectronic Engineering, Taiyuan University of Technology, Taiyuan 030024, China
}
\begin{abstract}
We perform a systematic QCD sum rule study of light tetraquark states with $J^{PC}=1^{--}$ in the diquark--antidiquark picture. A complete set of local interpolating currents is constructed and projected onto six flavor-isospin configurations ($q=u/d$): the isoscalar $q q\bar q\bar q$, $q s\bar q\bar s$, and $s s\bar s\bar s$ sectors, the isovector $q q\bar q\bar q$ and $q s\bar q\bar s$ sectors, and the isotensor $q q\bar q\bar q$ sector. The lowest masses in these sectors are derived to be $1.64^{+0.15}_{-0.14}$~GeV, $1.86^{+0.14}_{-0.14}$~GeV, $2.34^{+0.23}_{-0.30}$~GeV, $1.53^{+0.17}_{-0.19}$~GeV, $1.86^{+0.14}_{-0.14}$~GeV, and $2.24^{+0.12}_{-0.14}$~GeV, respectively. We further compare the present $1^{--}$ tetraquark spectrum with previous QCD sum rule results for the $1^{-+}$ tetraquark and hybrid states~\cite{Su:2025bhv}, aiming to provide useful information for distinguishing tetraquark and hybrid configurations in the light hadron spectrum. As an additional improvement, we complete the previously missing isotensor $1^{-+}$ tetraquark entry and obtain its lowest mass to be $M=2.19^{+0.26}_{-0.24}~\mathrm{GeV}$, which is included in the spectral comparison.
\end{abstract}

\maketitle
\pagenumbering{arabic}
%
\section{Introduction}
\label{sec:intro}

A hadron is a composite subatomic particle made of quarks and gluons and bound together by the strong interaction. In the traditional quark model, mesons are described as quark--antiquark bound states, while baryons are composed of three quarks. This simple classification scheme has achieved remarkable success in describing the observed hadron spectrum and in organizing the properties of conventional hadrons~\cite{pdg}. Nevertheless, quantum chromodynamics (QCD) also allows more complicated color-singlet configurations, such as tetraquark states, hadronic molecules, glueballs, and hybrid states~\cite{Klempt:2007cp,Crede:2008vw,Mathieu:2008me,Meyer:2010ku,Meyer:2015eta,Ochs:2013gi,Brambilla:2014jmp,Sonnenschein:2016pim,Briceno:2017max,Guo:2017jvc,Bass:2018xmz,Ketzer:2019wmd,Roberts:2021nhw,Fang:2021wes,Jin:2021vct,Gross:2022hyw,Chen:2022asf,Liu:2024uxn,Wang:2025dur,Luo:2025sns,Wang:2025sic,Dai:2026fkg}.

Among various quantum-number channels, the vector channel with $J^{PC}=1^{--}$ plays a special role. Although these quantum numbers are not exotic, vector states are experimentally important because they can be directly produced in electron--positron annihilation. The light vector meson spectrum is already rather rich in the conventional quark--antiquark picture, while additional multiquark components may also contribute in the same energy region. Therefore, a systematic study of light tetraquark states with $J^{PC}=1^{--}$ is useful for understanding the possible role of compact four-quark configurations in the light vector spectrum.

The present work is also motivated by the distinction between tetraquark and hybrid configurations. In particular, the $J^{PC}=1^{-+}$ channel has attracted considerable interest because these quantum numbers are forbidden for conventional quark--antiquark mesons. Experimentally, the isovector $\pi_1(1600)$ and the isoscalar $\eta_1(1855)$ are important candidates with exotic quantum numbers~\cite{JPAC:2018zyd,COMPASS:2021ogp,BESIII:2022riz}. Their possible internal structures have been discussed in various pictures, including hybrid states~\cite{Chen:2022isv,Qiu:2022ktc,Shastry:2022mhk,Chen:2023ukh,Esmer:2025xss}, tightly bound tetraquarks~\cite{Wan:2022xkx}, and hadronic molecules~\cite{Dong:2022cuw,Yang:2022rck,Liu:2024lph}. However, distinguishing a hybrid state from a compact tetraquark state remains a challenging problem, since the two configurations may appear in similar mass regions and may couple to similar hadronic channels.

For this reason, it is useful to investigate tetraquark and hybrid spectra in different quantum-number channels within a common theoretical framework. The $1^{-+}$ channel provides a direct probe of exotic quantum numbers, while the $1^{--}$ channel is experimentally accessible through electron--positron annihilation. A combined study of these two channels may therefore offer a broader perspective on the similarities and differences between tetraquark and hybrid configurations. In our previous work we studied the $1^{-+}$ tetraquark and hybrid spectra using QCD sum rules~\cite{Chen:2008qw,Chen:2008ne,Chen:2010ic,Huang:2010dc,Chen:2022qpd,Tan:2024grd,Su:2025bhv}. In the present work we extend this line of investigation to the light $1^{--}$ tetraquark sector, while the corresponding $1^{--}$ hybrid states will be investigated in future studies.

We systematically construct local diquark--antidiquark interpolating currents for light tetraquark states with $J^{PC}=1^{--}$. Their flavor structures are classified into the symmetric $\mathbf{6}_f\otimes\bar{\mathbf{6}}_f$, antisymmetric $\bar{\mathbf{3}}_f\otimes\mathbf{3}_f$, and mixed $(\bar{\mathbf{3}}_f\otimes\bar{\mathbf{6}}_f)\oplus(\mathbf{6}_f\otimes\mathbf{3}_f)$ representations. We then project these currents onto the isoscalar, isovector, and isotensor channels, and perform QCD sum rule analyses for the corresponding $q q\bar q\bar q$, $q s\bar q\bar s$, and $s s\bar s\bar s$ configurations. We also compare the resulting mass spectrum with our previous results for the $1^{-+}$ tetraquark and hybrid states~\cite{Su:2025bhv}, thereby providing a useful reference for future studies of light tetraquark and hybrid systems.

This paper is organized as follows. In Sec.~\ref{sec:current} we construct the $J^{PC}=1^{--}$ tetraquark interpolating currents and classify them according to their flavor and isospin structures. In Sec.~\ref{sec:sumrule} we formulate the QCD sum rules and present the relevant operator product expansion. In Sec.~\ref{sec:numeri} we perform the numerical analysis and extract the masses and decay constants. Finally, Sec.~\ref{sec:summary} summarizes the main results.

\section{Interpolating Currents}
\label{sec:current}

In this section we construct the local interpolating currents for the
$J^{PC}=1^{--}$ tetraquark states in the diquark--antidiquark picture.
Compared with conventional quark--antiquark mesons, tetraquark systems
possess much richer internal structures. It is therefore necessary to
organize the corresponding interpolating currents systematically in
terms of their flavor, color, and Lorentz structures, rather than
introducing them in an ad hoc manner. Such a classification is
particularly useful for the present analysis, since the isospin sectors
$I=0$, $1$, and $2$ will be discussed separately below.

We first summarize the notation used throughout this section. The
symbol $q$ denotes a light $u$ or $d$ quark, while $s$ denotes a
strange quark. The fields $\ell_{1a}$, $\ell_{2b}$,
$\bar{\ell}_{3c}$, and $\bar{\ell}_{4d}$ denote the quark and
antiquark fields entering the formal construction of the currents; the
corresponding flavor assignments will be specified explicitly once the
isospin channels are projected out. The indices $a,b,c,d$ are color
indices, the superscript $T$ denotes transpose in Dirac space, and $C$
is the charge-conjugation matrix. 

\subsection{General $(\ell\ell)(\bar{\ell}\bar{\ell})$ currents and symmetry decomposition}

The general local diquark--antidiquark current can be written as
\begin{equation}
J_\mu \sim \left(\ell_{1a}^T C \Gamma_1 \ell_{2b}\right)
\left(\bar{\ell}_{3c}\Gamma_{2\mu} C \bar{\ell}_{4d}^T\right),
\end{equation}
where the explicit form is determined by the requirement that the
current be a color singlet carrying the quantum numbers
$J^{PC}=1^{--}$.

\begin{figure*}[hbt]
\begin{center}
\scalebox{1}{\includegraphics{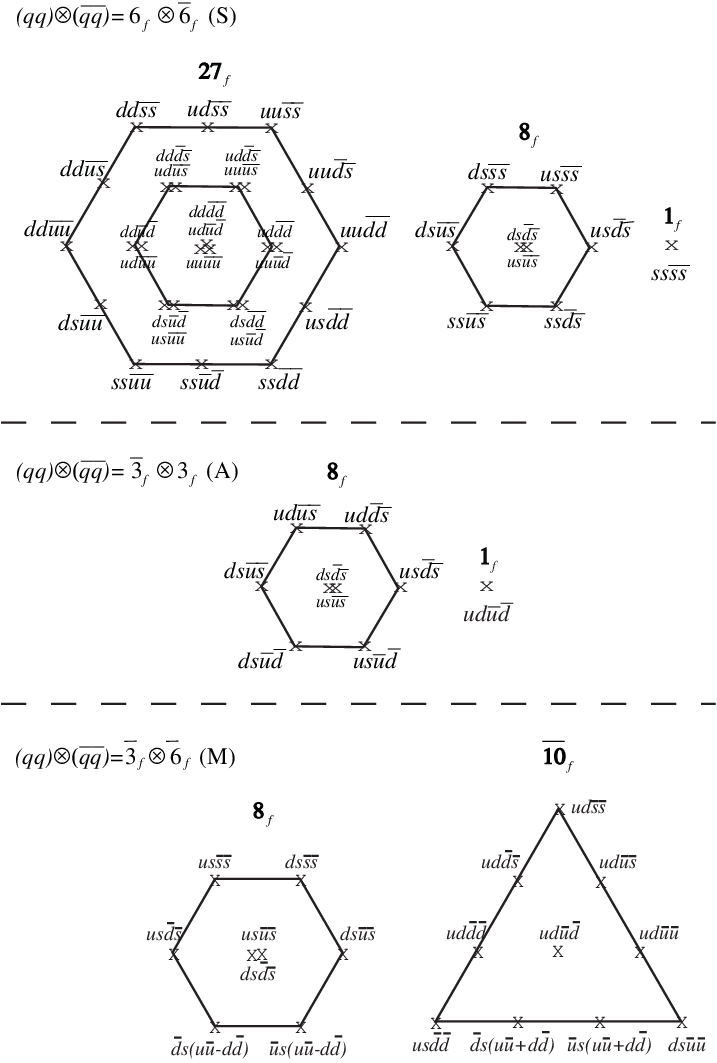}}
\caption{Weight diagrams for $\mathbf{6}_f\otimes\bar{\mathbf{6}}_f$
($\mathbf{S}$) (top panel),
$\bar{\mathbf{3}}_f\otimes\mathbf{3}_f$
($\mathbf{A}$) (middle panel), and
$\bar{\mathbf{3}}_f\otimes\bar{\mathbf{6}}_f$
($\mathbf{M}$) (bottom panel). The weight diagram for
$\mathbf{6}_f\otimes\mathbf{3}_f$ ($\mathbf{M}$) is obtained from the
bottom panel by charge conjugation.}
\label{fig:tetra}
\end{center}
\end{figure*}

To construct currents with definite charge-conjugation parity, one
should first classify the corresponding flavor structures. Under charge
conjugation, a diquark is transformed into an antidiquark while its
flavor symmetry is preserved. Therefore, the diquark and antidiquark
parts must be combined in compatible flavor representations. These flavor structures are illustrated in
Fig.~\ref{fig:tetra} through the corresponding $SU(3)$ weight
diagrams. The $\mathbf{S}$ and $\mathbf{A}$ flavor structures lead
directly to currents with definite charge-conjugation parity. By
contrast, neither $\bar{\mathbf{3}}_f \otimes \bar{\mathbf{6}}_f$ nor
$\mathbf{6}_f \otimes \mathbf{3}_f$ alone has a definite $C$ parity. A
definite $C$ parity can only be obtained from suitable linear
combinations of these two mixed representations. The tetraquark
currents are thus organized into the three classes
$\mathbf{S}$, $\mathbf{A}$, and $\mathbf{M}$. Their allowed isospin assignments can be summarized as
\begin{equation}\label{eq:flavor}
\begin{aligned}
\mathbf{S} \equiv \mathbf{6}_f \otimes \bar{\mathbf{6}}_f
:
&\begin{cases}
q q \bar q \bar q\,(\mathbf{S}), & I = 0,1,2,\\
q s \bar q \bar s\,(\mathbf{S}), & I = 0,1,\\
s \bar s s \bar s\,(\mathbf{S}), & I = 0,
\end{cases}
\\[0.8em]
\mathbf{A} \equiv \bar{\mathbf{3}}_f \otimes \mathbf{3}_f
:
&\begin{cases}
q q \bar q \bar q\,(\mathbf{A}), & I = 0,\\
q s \bar q \bar s\,(\mathbf{A}), & I = 0,1,
\end{cases}
\\[0.8em]
\mathbf{M} \equiv
\left(\bar{\mathbf{3}}_f \otimes \bar{\mathbf{6}}_f\right)
\oplus
\left(\mathbf{6}_f \otimes \mathbf{3}_f\right)
:
&\begin{cases}
q q \bar q \bar q\,(\mathbf{M}), & I = 1,\\
q s \bar q \bar s\,(\mathbf{M}), & I = 0,1.
\end{cases}
\end{aligned}
\end{equation}

We now list the $J^{PC}=1^{--}$ diquark--antidiquark currents
explicitly. For the symmetric flavor structure
$\mathbf{6}_f \otimes \bar{\mathbf{6}}_f$ ($\mathbf{S}$), there are two
independent $(\ell\ell)(\bar{\ell}\bar{\ell})$ currents:
\begin{align}
\psi^S_{1\mu}={}&
\ell_{1a}^T C \gamma_5 \ell_{2b}
\left(
\bar{\ell}_{3a}\gamma_\mu\gamma_5 C\bar{\ell}_{4b}^T
+\bar{\ell}_{3b}\gamma_\mu\gamma_5 C\bar{\ell}_{4a}^T
\right)
\nonumber\\
&-\ell_{1a}^T C \gamma_\mu\gamma_5 \ell_{2b}
\left(
\bar{\ell}_{3a}\gamma_5 C\bar{\ell}_{4b}^T
+\bar{\ell}_{3b}\gamma_5 C\bar{\ell}_{4a}^T
\right),
\label{eq:currentS1}
\\[0.4em]
\psi^S_{2\mu}={}&
\ell_{1a}^T C \gamma^\nu \ell_{2b}
\left(
\bar{\ell}_{3a}\sigma_{\mu\nu} C\bar{\ell}_{4b}^T
-\bar{\ell}_{3b}\sigma_{\mu\nu} C\bar{\ell}_{4a}^T
\right)
\nonumber\\
&-\ell_{1a}^T C \sigma_{\mu\nu} \ell_{2b}
\left(
\bar{\ell}_{3a}\gamma^\nu C\bar{\ell}_{4b}^T
-\bar{\ell}_{3b}\gamma^\nu C\bar{\ell}_{4a}^T
\right).
\label{eq:currentS2}
\end{align}

For the antisymmetric flavor structure
$\bar{\mathbf{3}}_f \otimes \mathbf{3}_f$ ($\mathbf{A}$), there are
likewise two independent currents:
\begin{align}
\psi^A_{1\mu}={}&
\ell_{1a}^T C \gamma_5 \ell_{2b}
\left(
\bar{\ell}_{3a}\gamma_\mu\gamma_5 C\bar{\ell}_{4b}^T
-\bar{\ell}_{3b}\gamma_\mu\gamma_5 C\bar{\ell}_{4a}^T
\right)
\nonumber\\
&-\ell_{1a}^T C \gamma_\mu\gamma_5 \ell_{2b}
\left(
\bar{\ell}_{3a}\gamma_5 C\bar{\ell}_{4b}^T
-\bar{\ell}_{3b}\gamma_5 C\bar{\ell}_{4a}^T
\right),
\label{eq:currentA1}
\\[0.4em]
\psi^A_{2\mu}={}&
\ell_{1a}^T C \gamma^\nu \ell_{2b}
\left(
\bar{\ell}_{3a}\sigma_{\mu\nu} C\bar{\ell}_{4b}^T
+\bar{\ell}_{3b}\sigma_{\mu\nu} C\bar{\ell}_{4a}^T
\right)
\nonumber\\
&-\ell_{1a}^T C \sigma_{\mu\nu} \ell_{2b}
\left(
\bar{\ell}_{3a}\gamma^\nu C\bar{\ell}_{4b}^T
+\bar{\ell}_{3b}\gamma^\nu C\bar{\ell}_{4a}^T
\right).
\label{eq:currentA2}
\end{align}

For the mixed flavor structure
$\left(\bar{\mathbf{3}}_f \otimes \bar{\mathbf{6}}_f\right)
\oplus
\left(\mathbf{6}_f \otimes \mathbf{3}_f\right)$, neither component
alone has a definite charge-conjugation parity. We therefore introduce
the currents $\psi^{ML}_{i\mu}$ in
$\bar{\mathbf{3}}_f \otimes \bar{\mathbf{6}}_f$ and the currents
$\psi^{MR}_{i\mu}$ in $\mathbf{6}_f \otimes \mathbf{3}_f$. The four independent $\psi^{ML}_{i\mu}$ currents are
\begin{align}
\psi^{ML}_{1\mu}={}&
\ell_{1a}^T C \gamma_\mu \ell_{2b}
\left(
\bar{\ell}_{3a} C\bar{\ell}_{4b}^T
+\bar{\ell}_{3b} C\bar{\ell}_{4a}^T
\right),
\nonumber \\
\psi^{ML}_{2\mu}={}&
\ell_{1a}^T C \sigma_{\mu\nu}\gamma_5 \ell_{2b}
\left(
\bar{\ell}_{3a}\gamma^\nu\gamma_5 C\bar{\ell}_{4b}^T
+\bar{\ell}_{3b}\gamma^\nu\gamma_5 C\bar{\ell}_{4a}^T
\right),
\nonumber 
\\
\psi^{ML}_{3\mu}={}&
\ell_{1a}^T C \ell_{2b}
\left(
\bar{\ell}_{3a}\gamma_\mu C\bar{\ell}_{4b}^T
-\bar{\ell}_{3b}\gamma_\mu C\bar{\ell}_{4a}^T
\right),
\nonumber \\
\psi^{ML}_{4\mu}={}&
\ell_{1a}^T C \gamma^\nu\gamma_5 \ell_{2b}
\left(
\bar{\ell}_{3a}\sigma_{\mu\nu}\gamma_5 C\bar{\ell}_{4b}^T
-\bar{\ell}_{3b}\sigma_{\mu\nu}\gamma_5 C\bar{\ell}_{4a}^T
\right).
\end{align}
Similarly, the four independent $\psi^{MR}_{i\mu}$ currents are
\begin{align}
\psi^{MR}_{1\mu}={}&
\ell_{1a}^T C \ell_{2b}
\left(
\bar{\ell}_{3a}\gamma_\mu C\bar{\ell}_{4b}^T
+\bar{\ell}_{3b}\gamma_\mu C\bar{\ell}_{4a}^T
\right),
\nonumber 
\\
\psi^{MR}_{2\mu}={}&
\ell_{1a}^T C \gamma^\nu\gamma_5 \ell_{2b}
\left(
\bar{\ell}_{3a}\sigma_{\mu\nu}\gamma_5 C\bar{\ell}_{4b}^T
+\bar{\ell}_{3b}\sigma_{\mu\nu}\gamma_5 C\bar{\ell}_{4a}^T
\right),
\nonumber 
\\
\psi^{MR}_{3\mu}={}&
\ell_{1a}^T C \gamma_\mu \ell_{2b}
\left(
\bar{\ell}_{3a} C\bar{\ell}_{4b}^T
-\bar{\ell}_{3b} C\bar{\ell}_{4a}^T
\right),
\nonumber 
\\
\psi^{MR}_{4\mu}={}&
\ell_{1a}^T C \sigma_{\mu\nu}\gamma_5 \ell_{2b}
\left(
\bar{\ell}_{3a}\gamma^\nu\gamma_5 C\bar{\ell}_{4b}^T
-\bar{\ell}_{3b}\gamma^\nu\gamma_5 C\bar{\ell}_{4a}^T
\right).
\end{align}
All these currents have $J^P=1^-$. The mixed currents do not
individually have definite charge-conjugation parity, but currents
with definite $C$ parity can be formed as
\begin{equation}\label{eq:currentM}
\psi^{M,\pm}_{i\mu}=\psi^{ML}_{i\mu}\pm\psi^{MR}_{i\mu},
\end{equation}
where the `$+$' and `$-$' combinations correspond to positive and
negative charge-conjugation parity, respectively. Since the present
work is concerned with the $J^{PC}=1^{--}$ tetraquark states, only the
negative-$C$ combinations will be used in the following analysis, and we shall simply denote them as $\psi^{M,-}_{i\mu} \to \psi^{M}_{i\mu}$.

The above construction provides the general
$(\ell\ell)(\bar{\ell}\bar{\ell})$ basis for the
$J^{PC}=1^{--}$ tetraquark currents. In the following subsections these currents are projected onto the isospin channels $I=0$, $1$, and $2$.

\subsection{Isoscalar ($I=0$) Currents}
\label{subsec:isocurrent0}

According to Eq.~\eqref{eq:flavor}, the isoscalar sector receives
contributions from the symmetric ($\mathbf{S}$), antisymmetric
($\mathbf{A}$), and mixed ($\mathbf{M}$) flavor representations. More
specifically, the $I=0$ currents arise from
$q q \bar q \bar q$, $q s \bar q \bar s$, and $s s \bar s \bar s$ in
the $\mathbf{S}$ representation, from $q q \bar q \bar q$ and
$q s \bar q \bar s$ in the $\mathbf{A}$ representation, and from
$q s \bar q \bar s$ only in the $\mathbf{M}$ representation. We denote
the explicit isoscalar currents by
$\psi^{R(f),\,I=0}_{i\mu}$, where $R=S,A,M$ labels the flavor
representation, $f=q q \bar q \bar q$, $q s \bar q \bar s$, or
$s s \bar s \bar s$ specifies the flavor content, and $i$ enumerates
the independent currents in each class.

For the symmetric flavor structure
$\mathbf{6}_f \otimes \bar{\mathbf{6}}_f$ ($\mathbf{S}$), there are six
independent isoscalar currents:
\begin{subequations}\label{eq:I0S}
\begin{align}
\psi^{S(q q \bar q \bar q),\,I=0}_{1\mu}
\equiv{}& \psi^S_{1\mu}(q q \bar q \bar q)
\label{eq:I0S1}\\
\sim{}&
u_a^T C \gamma_5 d_b
\left(
\bar{u}_a \gamma_\mu \gamma_5 C \bar{d}_b^T
+\bar{u}_b \gamma_\mu \gamma_5 C \bar{d}_a^T
\right)
\nonumber\\
&-
u_a^T C \gamma_\mu \gamma_5 d_b
\left(
\bar{u}_a \gamma_5 C \bar{d}_b^T
+\bar{u}_b \gamma_5 C \bar{d}_a^T
\right),
\nonumber\\[0.4em]
\psi^{S(q q \bar q \bar q),\,I=0}_{2\mu}
\equiv{}& \psi^S_{2\mu}(q q \bar q \bar q)
\label{eq:I0S2}\\
\sim{}&
u_a^T C \gamma^\nu d_b
\left(
\bar{u}_a \sigma_{\mu\nu} C \bar{d}_b^T
-\bar{u}_b \sigma_{\mu\nu} C \bar{d}_a^T
\right)
\nonumber\\
&-
u_a^T C \sigma_{\mu\nu} d_b
\left(
\bar{u}_a \gamma^\nu C \bar{d}_b^T
-\bar{u}_b \gamma^\nu C \bar{d}_a^T
\right),
\nonumber\\[0.6em]
\psi^{S(q s \bar q \bar s),\,I=0}_{1\mu}
\equiv{}& \psi^S_{1\mu}(q s \bar q \bar s)
\label{eq:I0S3}\\
\sim{}&
u_a^T C \gamma_5 s_b
\left(
\bar{u}_a \gamma_\mu \gamma_5 C \bar{s}_b^T
+\bar{u}_b \gamma_\mu \gamma_5 C \bar{s}_a^T
\right)
\nonumber\\
&-
u_a^T C \gamma_\mu \gamma_5 s_b
\left(
\bar{u}_a \gamma_5 C \bar{s}_b^T
+\bar{u}_b \gamma_5 C \bar{s}_a^T
\right),
\nonumber\\[0.4em]
\psi^{S(q s \bar q \bar s),\,I=0}_{2\mu}
\equiv{}& \psi^S_{2\mu}(q s \bar q \bar s)
\label{eq:I0S4}\\
\sim{}&
u_a^T C \gamma^\nu s_b
\left(
\bar{u}_a \sigma_{\mu\nu} C \bar{s}_b^T
-\bar{u}_b \sigma_{\mu\nu} C \bar{s}_a^T
\right)
\nonumber\\
&-
u_a^T C \sigma_{\mu\nu} s_b
\left(
\bar{u}_a \gamma^\nu C \bar{s}_b^T
-\bar{u}_b \gamma^\nu C \bar{s}_a^T
\right),
\nonumber
\\[0.6em]
\psi^{S(s s \bar s \bar s),\,I=0}_{1\mu}
\equiv{}& \psi^S_{1\mu}(s s \bar s \bar s)
\label{eq:I0S5}\\
={}&
s_a^T C \gamma_5 s_b
\left(
\bar{s}_a \gamma_\mu \gamma_5 C \bar{s}_b^T
+\bar{s}_b \gamma_\mu \gamma_5 C \bar{s}_a^T
\right)
\nonumber\\
&-
s_a^T C \gamma_\mu \gamma_5 s_b
\left(
\bar{s}_a \gamma_5 C \bar{s}_b^T
+\bar{s}_b \gamma_5 C \bar{s}_a^T
\right),
\nonumber\\[0.4em]
\psi^{S(s s \bar s \bar s),\,I=0}_{2\mu}
\equiv{}& \psi^S_{2\mu}(s s \bar s \bar s)
\label{eq:I0S6}\\
={}&
s_a^T C \gamma^\nu s_b
\left(
\bar{s}_a \sigma_{\mu\nu} C \bar{s}_b^T
-\bar{s}_b \sigma_{\mu\nu} C \bar{s}_a^T
\right)
\nonumber\\
&-
s_a^T C \sigma_{\mu\nu} s_b
\left(
\bar{s}_a \gamma^\nu C \bar{s}_b^T
-\bar{s}_b \gamma^\nu C \bar{s}_a^T
\right).
\nonumber
\end{align}
\end{subequations}
We use the symbol ``$\sim$'' to indicate that the quark contents written here are only schematic. For example, in the current $\psi^{S(q s \bar q \bar s),\,I=0}_{1\mu}$, the component $us\bar{u}\bar{s}$ alone does not have definite isospin $I=0$; the proper flavor structure should instead be the isoscalar combination $(us\bar{u}\bar{s}+ds\bar{d}\bar{s})$, i.e.,
\begin{align}
\psi^{S(q s \bar q \bar s),\,I=0}_{1\mu}
\equiv{}& \psi^S_{1\mu}(q s \bar q \bar s)
\\
={}&
u_a^T C \gamma_5 s_b
\left(
\bar{u}_a \gamma_\mu \gamma_5 C \bar{s}_b^T
+\bar{u}_b \gamma_\mu \gamma_5 C \bar{s}_a^T
\right)
\nonumber\\
&-
u_a^T C \gamma_\mu \gamma_5 s_b
\left(
\bar{u}_a \gamma_5 C \bar{s}_b^T
+\bar{u}_b \gamma_5 C \bar{s}_a^T
\right)
\nonumber\\
+&
d_a^T C \gamma_5 s_b
\left(
\bar{d}_a \gamma_\mu \gamma_5 C \bar{s}_b^T
+\bar{d}_b \gamma_\mu \gamma_5 C \bar{s}_a^T
\right)
\nonumber\\
&-
d_a^T C \gamma_\mu \gamma_5 s_b
\left(
\bar{d}_a \gamma_5 C \bar{s}_b^T
+\bar{d}_b \gamma_5 C \bar{s}_a^T
\right) \, .
\nonumber
\end{align}
However, in the following QCD sum rule analysis, we neglect the up- and down-quark masses and assume $\langle \bar{u}u\rangle=\langle \bar{d}d\rangle$. Under these approximations, the QCD sum rule results obtained from $\psi^{S(q s \bar q \bar s),\,I=0}_{1\mu}$ with quark content $us\bar{u}\bar{s}$ are identical to those obtained from the more precise flavor combination $(us\bar{u}\bar{s}+ds\bar{d}\bar{s})$. The full isospin multiplets can be reconstructed straightforwardly using the standard isospin formalism.

For the antisymmetric flavor structure
$\bar{\mathbf{3}}_f \otimes \mathbf{3}_f$ ($\mathbf{A}$), there are
four independent isoscalar currents:
\begin{subequations}\label{eq:I0A}
\begin{align}
\psi^{A(q q \bar q \bar q),\,I=0}_{1\mu}
\equiv{}& \psi^A_{1\mu}(q q \bar q \bar q)
\label{eq:I0A1}\\
\sim{}&
u_a^T C \gamma_5 d_b
\left(
\bar{u}_a \gamma_\mu \gamma_5 C \bar{d}_b^T
-\bar{u}_b \gamma_\mu \gamma_5 C \bar{d}_a^T
\right)
\nonumber\\
&-
u_a^T C \gamma_\mu \gamma_5 d_b
\left(
\bar{u}_a \gamma_5 C \bar{d}_b^T
-\bar{u}_b \gamma_5 C \bar{d}_a^T
\right),
\nonumber\\[0.4em]
\psi^{A(q q \bar q \bar q),\,I=0}_{2\mu}
\equiv{}& \psi^A_{2\mu}(q q \bar q \bar q)
\label{eq:I0A2}\\
\sim{}&
u_a^T C \gamma^\nu d_b
\left(
\bar{u}_a \sigma_{\mu\nu} C \bar{d}_b^T
+\bar{u}_b \sigma_{\mu\nu} C \bar{d}_a^T
\right)
\nonumber\\
&-
u_a^T C \sigma_{\mu\nu} d_b
\left(
\bar{u}_a \gamma^\nu C \bar{d}_b^T
+\bar{u}_b \gamma^\nu C \bar{d}_a^T
\right),
\nonumber\\[0.6em]
\psi^{A(q s \bar q \bar s),\,I=0}_{1\mu}
\equiv{}& \psi^A_{1\mu}(q s \bar q \bar s)
\label{eq:I0A3}\\
\sim{}&
u_a^T C \gamma_5 s_b
\left(
\bar{u}_a \gamma_\mu \gamma_5 C \bar{s}_b^T
-\bar{u}_b \gamma_\mu \gamma_5 C \bar{s}_a^T
\right)
\nonumber\\
&-
u_a^T C \gamma_\mu \gamma_5 s_b
\left(
\bar{u}_a \gamma_5 C \bar{s}_b^T
-\bar{u}_b \gamma_5 C \bar{s}_a^T
\right),
\nonumber\\[0.4em]
\psi^{A(q s \bar q \bar s),\,I=0}_{2\mu}
\equiv{}& \psi^A_{2\mu}(q s \bar q \bar s)
\label{eq:I0A4}\\
\sim{}&
u_a^T C \gamma^\nu s_b
\left(
\bar{u}_a \sigma_{\mu\nu} C \bar{s}_b^T
+\bar{u}_b \sigma_{\mu\nu} C \bar{s}_a^T
\right)
\nonumber\\
&-
u_a^T C \sigma_{\mu\nu} s_b
\left(
\bar{u}_a \gamma^\nu C \bar{s}_b^T
+\bar{u}_b \gamma^\nu C \bar{s}_a^T
\right).
\nonumber
\end{align}
\end{subequations}

For the mixed flavor structure
$\left(\bar{\mathbf{3}}_f \otimes \bar{\mathbf{6}}_f\right)
\oplus
\left(\mathbf{6}_f \otimes \mathbf{3}_f\right)$ ($\mathbf{M}$), the
isoscalar sector is contributed by the $q s \bar q \bar s$ channel
only. The corresponding four independent currents are
\begin{subequations}\label{eq:I0M}
\begin{align}
\psi&^{M(q s \bar q \bar s),\,I=0}_{1\mu}
\equiv{} \psi^M_{1\mu}(q s \bar q \bar s)
\label{eq:I0M1}\\
&\sim{}
u_a^T C \gamma_\mu s_b
\left(
\bar{u}_a C \bar{s}_b^T
+\bar{u}_b C \bar{s}_a^T
\right)
\nonumber\\
&-
u_a^T C s_b
\left(
\bar{u}_a \gamma_\mu C \bar{s}_b^T
+\bar{u}_b \gamma_\mu C \bar{s}_a^T
\right),
\nonumber\\[0.4em]
\psi&^{M(q s \bar q \bar s),\,I=0}_{2\mu}
\equiv{} \psi^M_{2\mu}(q s \bar q \bar s)
\label{eq:I0M2}\\
&\sim{}
u_a^T C \sigma_{\mu\nu}\gamma_5 s_b
\left(
\bar{u}_a \gamma^\nu\gamma_5 C \bar{s}_b^T
+\bar{u}_b \gamma^\nu\gamma_5 C \bar{s}_a^T
\right)
\nonumber\\
&-
u_a^T C \gamma^\nu\gamma_5 s_b
\left(
\bar{u}_a \sigma_{\mu\nu}\gamma_5 C \bar{s}_b^T
+\bar{u}_b \sigma_{\mu\nu}\gamma_5 C \bar{s}_a^T
\right),
\nonumber\\[0.4em]
\psi&^{M(q s \bar q \bar s),\,I=0}_{3\mu}
\equiv{} \psi^M_{3\mu}(q s \bar q \bar s)
\label{eq:I0M3}\\
&\sim{}
u_a^T C s_b
\left(
\bar{u}_a \gamma_\mu C \bar{s}_b^T
-\bar{u}_b \gamma_\mu C \bar{s}_a^T
\right)
\nonumber\\
&-
u_a^T C \gamma_\mu s_b
\left(
\bar{u}_a C \bar{s}_b^T
-\bar{u}_b C \bar{s}_a^T
\right),
\nonumber\\[0.4em]
\psi&^{M(q s \bar q \bar s),\,I=0}_{4\mu}
\equiv{} \psi^M_{4\mu}(q s \bar q \bar s)
\label{eq:I0M4}\\
&\sim{}
u_a^T C \gamma^\nu\gamma_5 s_b
\left(
\bar{u}_a \sigma_{\mu\nu}\gamma_5 C \bar{s}_b^T
-\bar{u}_b \sigma_{\mu\nu}\gamma_5 C \bar{s}_a^T
\right)
\nonumber\\
&-
u_a^T C \sigma_{\mu\nu}\gamma_5 s_b
\left(
\bar{u}_a \gamma^\nu\gamma_5 C \bar{s}_b^T
-\bar{u}_b \gamma^\nu\gamma_5 C \bar{s}_a^T
\right).
\nonumber
\end{align}
\end{subequations}

\subsection{Isovector ($I=1$) Currents}
\label{subsec:isocurrent1}

According to Eq.~\eqref{eq:flavor}, the isovector sector receives
contributions from the symmetric ($\mathbf{S}$), antisymmetric
($\mathbf{A}$), and mixed ($\mathbf{M}$) flavor representations. More
specifically, the $I=1$ currents arise from
$q q \bar q \bar q$ and $q s \bar q \bar s$ in the $\mathbf{S}$
representation, from $q s \bar q \bar s$ in the $\mathbf{A}$
representation, and from $q q \bar q \bar q$ as well as
$q s \bar q \bar s$ in the $\mathbf{M}$ representation. We denote the
explicit isovector currents by $\psi^{R(f),\,I=1}_{i\mu}$.

For the symmetric flavor structure
$\mathbf{6}_f \otimes \bar{\mathbf{6}}_f$ ($\mathbf{S}$), there are
four independent isovector currents:
\begin{subequations}\label{eq:I1S}
\begin{align}
\psi^{S(q q \bar q \bar q),\,I=1}_{1\mu}
\equiv{}& \psi^S_{1\mu}(q q \bar q \bar q)
\label{eq:I1S1}\\
\sim{}&
u_a^T C \gamma_5 d_b
\left(
\bar{u}_a \gamma_\mu\gamma_5 C \bar{d}_b^T
+\bar{u}_b \gamma_\mu\gamma_5 C \bar{d}_a^T
\right)
\nonumber\\
&-
u_a^T C \gamma_\mu\gamma_5 d_b
\left(
\bar{u}_a \gamma_5 C \bar{d}_b^T
+\bar{u}_b \gamma_5 C \bar{d}_a^T
\right),
\nonumber
\\[0.4em]
\psi^{S(q q \bar q \bar q),\,I=1}_{2\mu}
\equiv{}& \psi^S_{2\mu}(q q \bar q \bar q)
\label{eq:I1S2}\\
\sim{}&
u_a^T C \gamma^\nu d_b
\left(
\bar{u}_a \sigma_{\mu\nu} C \bar{d}_b^T
-\bar{u}_b \sigma_{\mu\nu} C \bar{d}_a^T
\right)
\nonumber\\
&-
u_a^T C \sigma_{\mu\nu} d_b
\left(
\bar{u}_a \gamma^\nu C \bar{d}_b^T
-\bar{u}_b \gamma^\nu C \bar{d}_a^T
\right),
\nonumber
\\[0.6em]
\psi^{S(q s \bar q \bar s),\,I=1}_{1\mu}
\equiv{}& \psi^S_{1\mu}(q s \bar q \bar s)
\label{eq:I1S3}\\
\sim{}&
u_a^T C \gamma_5 s_b
\left(
\bar{u}_a \gamma_\mu\gamma_5 C \bar{s}_b^T
+\bar{u}_b \gamma_\mu\gamma_5 C \bar{s}_a^T
\right)
\nonumber\\
&-
u_a^T C \gamma_\mu\gamma_5 s_b
\left(
\bar{u}_a \gamma_5 C \bar{s}_b^T
+\bar{u}_b \gamma_5 C \bar{s}_a^T
\right),
\nonumber
\\[0.4em]
\psi^{S(q s \bar q \bar s),\,I=1}_{2\mu}
\equiv{}& \psi^S_{2\mu}(q s \bar q \bar s)
\label{eq:I1S4}\\
\sim{}&
u_a^T C \gamma^\nu s_b
\left(
\bar{u}_a \sigma_{\mu\nu} C \bar{s}_b^T
-\bar{u}_b \sigma_{\mu\nu} C \bar{s}_a^T
\right)
\nonumber\\
&-
u_a^T C \sigma_{\mu\nu} s_b
\left(
\bar{u}_a \gamma^\nu C \bar{s}_b^T
-\bar{u}_b \gamma^\nu C \bar{s}_a^T
\right).
\nonumber
\end{align}
\end{subequations}
It should be noted that the structure of the current $\psi^{S(q q \bar q \bar q),\,I=1}_{1\mu}$ in Eq.~(\ref{eq:I1S1}) is identical to that of the current $\psi^{S(q q \bar q \bar q),\,I=0}_{1\mu}$ in Eq.~(\ref{eq:I0S1}), although their exact expressions are not strictly identical once the proper flavor combinations are taken into account. Nevertheless, the corresponding QCD sum rule results turn out to be the same.

For the antisymmetric flavor structure
$\bar{\mathbf{3}}_f \otimes \mathbf{3}_f$ ($\mathbf{A}$), there are two
independent isovector currents:
\begin{subequations}\label{eq:I1A}
\begin{align}
\psi^{A(q s \bar q \bar s),\,I=1}_{1\mu}
\equiv{}& \psi^A_{1\mu}(q s \bar q \bar s)
\label{eq:I1A1}\\
\sim{}&
u_a^T C \gamma_5 s_b
\left(
\bar{u}_a \gamma_\mu\gamma_5 C \bar{s}_b^T
-\bar{u}_b \gamma_\mu\gamma_5 C \bar{s}_a^T
\right)
\nonumber\\
&-
u_a^T C \gamma_\mu\gamma_5 s_b
\left(
\bar{u}_a \gamma_5 C \bar{s}_b^T
-\bar{u}_b \gamma_5 C \bar{s}_a^T
\right),
\nonumber
\\[0.4em]
\psi^{A(q s \bar q \bar s),\,I=1}_{2\mu}
\equiv{}& \psi^A_{2\mu}(q s \bar q \bar s)
\label{eq:I1A2}\\
\sim{}&
u_a^T C \gamma^\nu s_b
\left(
\bar{u}_a \sigma_{\mu\nu} C \bar{s}_b^T
+\bar{u}_b \sigma_{\mu\nu} C \bar{s}_a^T
\right)
\nonumber\\
&-
u_a^T C \sigma_{\mu\nu} s_b
\left(
\bar{u}_a \gamma^\nu C \bar{s}_b^T
+\bar{u}_b \gamma^\nu C \bar{s}_a^T
\right).
\nonumber
\end{align}
\end{subequations}

For the mixed flavor structure
$\left(\bar{\mathbf{3}}_f \otimes \bar{\mathbf{6}}_f\right)
\oplus
\left(\mathbf{6}_f \otimes \mathbf{3}_f\right)$ ($\mathbf{M}$), the
isovector sector receives contributions from both the
$q q \bar q \bar q$ and $q s \bar q \bar s$ channels. The corresponding
independent currents are
\begin{subequations}\label{eq:I1M}
\begin{align}
\psi&^{M(q q \bar q \bar q),\,I=1}_{1\mu}
\equiv{} \psi^M_{1\mu}(q q \bar q \bar q)
\label{eq:I1M1}\\
&\sim{}
u_a^T C \gamma_\mu d_b
\left(
\bar{u}_a C \bar{d}_b^T
+\bar{u}_b C \bar{d}_a^T
\right)
\nonumber\\
&-
u_a^T C d_b
\left(
\bar{u}_a \gamma_\mu C \bar{d}_b^T
+\bar{u}_b \gamma_\mu C \bar{d}_a^T
\right),
\nonumber
\\[0.4em]
\psi&^{M(q q \bar q \bar q),\,I=1}_{2\mu}
\equiv{} \psi^M_{2\mu}(q q \bar q \bar q)
\label{eq:I1M2}\\
&\sim{}
u_a^T C \sigma_{\mu\nu}\gamma_5 d_b
\left(
\bar{u}_a \gamma^\nu\gamma_5 C \bar{d}_b^T
+\bar{u}_b \gamma^\nu\gamma_5 C \bar{d}_a^T
\right)
\nonumber\\
&-
u_a^T C \gamma^\nu\gamma_5 d_b
\left(
\bar{u}_a \sigma_{\mu\nu}\gamma_5 C \bar{d}_b^T
+\bar{u}_b \sigma_{\mu\nu}\gamma_5 C \bar{d}_a^T
\right),
\nonumber
\\[0.4em]
\psi&^{M(q q \bar q \bar q),\,I=1}_{3\mu}
\equiv{} \psi^M_{3\mu}(q q \bar q \bar q)
\label{eq:I1M3}\\
&\sim{}
u_a^T C d_b
\left(
\bar{u}_a \gamma_\mu C \bar{d}_b^T
-\bar{u}_b \gamma_\mu C \bar{d}_a^T
\right)
\nonumber\\
&-
u_a^T C \gamma_\mu d_b
\left(
\bar{u}_a C \bar{d}_b^T
-\bar{u}_b C \bar{d}_a^T
\right),
\nonumber
\\[0.4em]
\psi&^{M(q q \bar q \bar q),\,I=1}_{4\mu}
\equiv{} \psi^M_{4\mu}(q q \bar q \bar q)
\label{eq:I1M4}\\
&\sim{}
u_a^T C \gamma^\nu\gamma_5 d_b
\left(
\bar{u}_a \sigma_{\mu\nu}\gamma_5 C \bar{d}_b^T
-\bar{u}_b \sigma_{\mu\nu}\gamma_5 C \bar{d}_a^T
\right)
\nonumber\\
&-
u_a^T C \sigma_{\mu\nu}\gamma_5 d_b
\left(
\bar{u}_a \gamma^\nu\gamma_5 C \bar{d}_b^T
-\bar{u}_b \gamma^\nu\gamma_5 C \bar{d}_a^T
\right),
\nonumber
\\[0.6em]
\psi&^{M(q s \bar q \bar s),\,I=1}_{1\mu}
\equiv{} \psi^M_{1\mu}(q s \bar q \bar s)
\label{eq:I1M5}\\
&\sim{}
u_a^T C \gamma_\mu s_b
\left(
\bar{u}_a C \bar{s}_b^T
+\bar{u}_b C \bar{s}_a^T
\right)
\nonumber\\
&-
u_a^T C s_b
\left(
\bar{u}_a \gamma_\mu C \bar{s}_b^T
+\bar{u}_b \gamma_\mu C \bar{s}_a^T
\right),
\nonumber
\\[0.4em]
\psi&^{M(q s \bar q \bar s),\,I=1}_{2\mu}
\equiv{} \psi^M_{2\mu}(q s \bar q \bar s)
\label{eq:I1M6}\\
&\sim{}
u_a^T C \sigma_{\mu\nu}\gamma_5 s_b
\left(
\bar{u}_a \gamma^\nu\gamma_5 C \bar{s}_b^T
+\bar{u}_b \gamma^\nu\gamma_5 C \bar{s}_a^T
\right)
\nonumber\\
&-
u_a^T C \gamma^\nu\gamma_5 s_b
\left(
\bar{u}_a \sigma_{\mu\nu}\gamma_5 C \bar{s}_b^T
+\bar{u}_b \sigma_{\mu\nu}\gamma_5 C \bar{s}_a^T
\right),
\nonumber
\\[0.4em]
\psi&^{M(q s \bar q \bar s),\,I=1}_{3\mu}
\equiv{} \psi^M_{3\mu}(q s \bar q \bar s)
\label{eq:I1M7}\\
&\sim{}
u_a^T C s_b
\left(
\bar{u}_a \gamma_\mu C \bar{s}_b^T
-\bar{u}_b \gamma_\mu C \bar{s}_a^T
\right)
\nonumber\\
&-
u_a^T C \gamma_\mu s_b
\left(
\bar{u}_a C \bar{s}_b^T
-\bar{u}_b C \bar{s}_a^T
\right),
\nonumber
\\[0.4em]
\psi&^{M(q s \bar q \bar s),\,I=1}_{4\mu}
\equiv{} \psi^M_{4\mu}(q s \bar q \bar s)
\label{eq:I1M8}\\
&\sim{}
u_a^T C \gamma^\nu\gamma_5 s_b
\left(
\bar{u}_a \sigma_{\mu\nu}\gamma_5 C \bar{s}_b^T
-\bar{u}_b \sigma_{\mu\nu}\gamma_5 C \bar{s}_a^T
\right)
\nonumber\\
&-
u_a^T C \sigma_{\mu\nu}\gamma_5 s_b
\left(
\bar{u}_a \gamma^\nu\gamma_5 C \bar{s}_b^T
-\bar{u}_b \gamma^\nu\gamma_5 C \bar{s}_a^T
\right).
\nonumber
\end{align}
\end{subequations}

\subsection{Isotensor ($I=2$) Currents}
\label{subsec:isocurrent2}

According to Eq.~\eqref{eq:flavor}, the isotensor sector receives a
contribution only from the symmetric flavor representation
$\mathbf{6}_f \otimes \bar{\mathbf{6}}_f$ ($\mathbf{S}$), and only
through the $q q \bar q \bar q$ channel. There are two independent
currents, which we denote by
$\psi^{S(q q \bar q \bar q),\,I=2}_{i\mu}$:
\begin{subequations}\label{eq:I2S}
\begin{align}
\psi^{S(q q \bar q \bar q),\,I=2}_{1\mu}
\equiv{}& \psi^S_{1\mu}(q q \bar q \bar q)
\label{eq:I2S1}\\
\sim{}&
u_a^T C \gamma_5 d_b
\left(
\bar{u}_a \gamma_\mu\gamma_5 C \bar{d}_b^T
+\bar{u}_b \gamma_\mu\gamma_5 C \bar{d}_a^T
\right)
\nonumber\\
&-
u_a^T C \gamma_\mu\gamma_5 d_b
\left(
\bar{u}_a \gamma_5 C \bar{d}_b^T
+\bar{u}_b \gamma_5 C \bar{d}_a^T
\right),
\nonumber
\\[0.4em]
\psi^{S(q q \bar q \bar q),\,I=2}_{2\mu}
\equiv{}& \psi^S_{2\mu}(q q \bar q \bar q)
\label{eq:I2S2}\\
\sim{}&
u_a^T C \gamma^\nu d_b
\left(
\bar{u}_a \sigma_{\mu\nu} C \bar{d}_b^T
-\bar{u}_b \sigma_{\mu\nu} C \bar{d}_a^T
\right)
\nonumber\\
&-
u_a^T C \sigma_{\mu\nu} d_b
\left(
\bar{u}_a \gamma^\nu C \bar{d}_b^T
-\bar{u}_b \gamma^\nu C \bar{d}_a^T
\right).
\nonumber
\end{align}
\end{subequations}
It is worth noting once again that the structure of the current $\psi^{S(q q \bar q \bar q),\,I=2}_{1\mu}$ in Eq.~(\ref{eq:I2S1}) is identical to those of the currents $\psi^{S(q q \bar q \bar q),\,I=1}_{1\mu}$ in Eq.~(\ref{eq:I1S1}) and $\psi^{S(q q \bar q \bar q),\,I=0}_{1\mu}$ in Eq.~(\ref{eq:I0S1}), although their exact expressions are not strictly identical. Therefore, the corresponding QCD sum rule results turn out to be the same.

\section{QCD sum rules}
\label{sec:sumrule}

Over the past several decades, the QCD sum rule method has proven itself to be a powerful and effective non-perturbative technique~\cite{Shifman:1978bx,Reinders:1984sr,Lian:2026uys}. In sum rule analyses, we focus on two-point correlation functions, which are defined as:
\begin{equation}
\Pi_{\mu\nu}(q^2) \equiv i \int d^4x \, e^{iqx} \langle 0 | T \left( \eta_\mu(x) \, \eta_\nu^\dagger(0) \right) | 0 \rangle \, , \label{def:pi}
\end{equation}
where $\eta_\mu$ represents the interpolating current associated with the tetraquark state, denoted as $Y$. The Lorentz structure of the correlation function can be simplified to:
\begin{equation}
\Pi_{\mu\nu}(q^2) = \left( \frac{q_\mu q_\nu}{q^2} - g_{\mu\nu} \right) \Pi^{(1)}(q^2) + \frac{q_\mu q_\nu}{q^2} \Pi^{(0)}(q^2) \, . \label{def:pi1}
\end{equation}
We compute $\Pi^{(1)}(q^2)$ within the framework of the operator product expansion (OPE) in QCD, truncating the expansion at a certain order to extract the OPE spectral density $\rho_{\rm OPE}(s)$. This result is then matched with a hadronic parametrization to extract information about hadron properties.

At the hadronic level, the correlation function is represented as a dispersion relation involving a phenomenological spectral function:
\begin{equation}
\Pi^{(1)}(q^2) = \int_{s_<}^{\infty} \frac{\rho_{\rm phen}(s)}{s - q^2 - i\varepsilon} \, ds \, , \label{eq:disper}
\end{equation}
where $s_<$ is the physical threshold. The spectral density $\rho_{\rm phen}(s)$ is defined as:
\begin{eqnarray}
&& \rho_{\rm phen}(s) \times \left( \frac{q_\mu q_\nu}{q^2} - g_{\mu\nu} \right) 
\nonumber\\ & \equiv & \sum_n \delta(s - M_n^2) \langle 0 | \eta_\mu | n \rangle \langle n | \eta^\dagger_\nu | 0 \rangle \nonumber\\
& = & f_Y^2 \delta(s - M_Y^2) \epsilon_\mu \epsilon_\nu^* + \text{continuum} \, , \label{eq:rho}
\end{eqnarray}
where the coupling constant \( f_Y \) is defined via the matrix element
\begin{equation}
\langle 0 | \eta_\mu | Y \rangle = f_Y \, \epsilon_\mu \, ,
\label{coupling1}
\end{equation}
with \( \epsilon_\mu \) being the polarization vector.

In the second equation, we adopt the standard parametrization, assuming one-pole dominance for the ground state $Y$ and including a continuum contribution. The sum rule analysis is then performed after applying the Borel transformation to Eq.~(\ref{eq:disper}):
\begin{equation}
\Pi(\infty, M_B^2) \equiv \mathcal{B}_{M_B^2} \Pi^{(1)}(p^2) = \int_{s_<}^{\infty} e^{-s/M_B^2} \rho_{\rm phen}(s) \, ds \, . \label{eq:borel}
\end{equation}
Assuming that the contribution from the continuum states can be effectively approximated by the OPE spectral density above a threshold value $s_0$, we obtain the following sum rule equation:
\begin{equation}
\Pi(s_0, M_B^2) = f_Y^2 e^{-M_Y^2 / M_B^2} = \int_{s_<}^{s_0} e^{-s/M_B^2} \rho_{\rm OPE}(s) \, ds \, . \label{eq:fin}
\end{equation}
By differentiating this equation with respect to $1 / M_B^2$, we obtain the expression for the mass squared of the ground state $Y$:
\begin{equation}
M_Y^2 = \frac{\frac{\partial}{\partial(-1/M_B^2)} \Pi(s_0, M_B^2)}{\Pi(s_0, M_B^2)} = \frac{\int_{s_<}^{s_0} e^{-s/M_B^2} s \rho_{\rm OPE}(s) \, ds}{\int_{s_<}^{s_0} e^{-s/M_B^2} \rho_{\rm OPE}(s) \, ds} \, . \label{eq:LSR}
\end{equation}
Additionally, we arrive at the corollary:
\begin{equation}
f_Y^2 = \Pi(s_0, M_B^2) \times e^{M_Y^2 / M_B^2} \, . 
\label{eq:dec}
\end{equation}
In the next section, we will examine both Eqs.~(\ref{eq:LSR}) and (\ref{eq:dec}) as functions of parameters such as the Borel mass $M_B$ and the threshold value $s_0$, for various tetraquark currents.

We have performed the OPE calculation up to dimension ten. Here, we present the result for the current $\psi^M_{2\mu}(q s \bar q \bar s)$ defined in Eq.~(\ref{eq:I1M6}), while the results for other currents are provided in Appendix~\ref{app:ope}:

\begin{eqnarray}\nonumber
&&\Pi^{M(q s \bar q \bar s)}_{2\mu}(M_B^2)  =\\ \nonumber &&\int^{s_0}_{4m_s^2} \Bigg [
{ 1 \over 6144 \pi^6 } s^4
- { 3 m_s^2 \over 512 \pi^6 } s^3
+ \Big (
{ 11 \langle g_s^2 G G \rangle \over 18432 \pi^6 }
+ { 9 m_s^4 \over 256 \pi^6 } 
\\ \nonumber && + { m_s \langle \bar s s \rangle \over 32 \pi^4 }
\Big ) s^2 
+ \Big (
{ 7 m_s^2 \langle g_s^2 G G \rangle \over 6144 \pi^6 }
- { m_s^3 \langle \bar s s \rangle \over 4 \pi^4 }
\\ \nonumber && + { m_s \langle g_s \bar q \sigma G q \rangle \over 96 \pi^4 }
- { m_s \langle g_s \bar s \sigma G s \rangle \over 64 \pi^4 }
+ { \langle \bar q q \rangle^2 \over 12 \pi^2 }
+ { \langle \bar s s \rangle^2 \over 12 \pi^2 }
\Big ) s
\\ \nonumber && + \Big (
{ m_s \langle g_s^2 G G \rangle \langle \bar q q \rangle \over 64 \pi^4 }
- { 5 m_s \langle g_s^2 G G \rangle \langle \bar s s \rangle \over 256 \pi^4 }
- { 3 m_s^2 \langle \bar q q \rangle^2 \over 2 \pi^2 }
\\ \nonumber && + { m_s^2 \langle \bar s s \rangle^2 \over 8 \pi^2 }
+ { \langle \bar q q \rangle \langle g_s \bar q \sigma G q \rangle \over 16 \pi^2 }
- { \langle \bar q q \rangle \langle g_s \bar s \sigma G s \rangle \over 24 \pi^2 }
\\ \nonumber && - { \langle \bar s s \rangle \langle g_s \bar q \sigma G q \rangle \over 24 \pi^2 }
+ { \langle \bar s s \rangle \langle g_s \bar s \sigma G s \rangle \over 16 \pi^2 }
\Big )
\Bigg ] e^{-s/M_B^2} ds
\\ \nonumber && + \Big (
{ 5 m_s^3 \langle g_s^2 G G \rangle \langle \bar s s \rangle \over 1152 \pi^2 }
+ { 5 m_s \langle g_s^2 G G \rangle \langle g_s \bar q \sigma G q \rangle \over 1152 \pi^4 }
\\ \nonumber && - { 25 m_s \langle g_s^2 G G \rangle \langle g_s \bar s \sigma G s \rangle \over 4608 \pi^4 }
+ { 25 \langle g_s^2 G G \rangle \langle \bar q q \rangle^2 \over 1728 \pi^2 }
\\ \nonumber && - { 5 \langle g_s^2 G G \rangle \langle \bar q q \rangle \langle \bar s s \rangle \over 216 \pi^2 }
+ { 25 \langle g_s^2 G G \rangle \langle \bar s s \rangle^2 \over 1728 \pi^2 }
+ { m_s^4 \langle \bar q q \rangle^2 \over 2 \pi^2 }
\\ \nonumber && - { 7 m_s^2 \langle \bar q q \rangle \langle g_s \bar q \sigma G q \rangle \over 8 \pi^2 }
+ { m_s^2 \langle \bar q q \rangle \langle g_s \bar s \sigma G s \rangle \over 12 \pi^2 }
\\ \nonumber && + { 10 m_s \langle \bar q q \rangle^2 \langle \bar s s \rangle \over 3 }
- { \langle g_s \bar q \sigma G q \rangle \langle g_s \bar s \sigma G s \rangle \over 24 \pi^2 }
\Big ),
\end{eqnarray}

In the above equation, $\langle \bar{q}q \rangle$ represents the $D=3$ up/down quark condensate; $\langle g_s^2 GG \rangle$ refers to the $D=4$ gluon condensate; and $\langle g_s \bar{q} \sigma G q \rangle$ is the $D=5$ mixed condensate. As is customary, we assume vacuum saturation for higher-dimensional condensates, such as $\langle 0 | \bar q q \bar q q | 0 \rangle \sim \langle 0 | \bar q q | 0 \rangle \langle 0 | \bar q q | 0 \rangle$.

\section{Numerical Analysis}
\label{sec:numeri}

To reliably extract the mass using Eq.~(\ref{eq:LSR}), it is essential to determine the appropriate regions for the two free parameters: the threshold value \( s_0 \) and the Borel mass \( M_B \). We present the result for the current $\psi^M_{2\mu}(q s \bar q \bar s)$ defined in Eq.~(\ref{eq:I1M6}), and we select the following values for various condensates and \( m_s \) at 2 GeV~\cite{pdg,Yang:1993bp,Gimenez:2005nt,Jamin:2002ev,Ioffe:2002be,Ovchinnikov:1988gk,Narison:2018dcr}:
\begin{eqnarray}
\nonumber && \langle \bar q q \rangle = -(0.240 \pm 0.010)\, \mathrm{GeV}^3 \,,
\\
\nonumber && \langle \bar s s \rangle = (0.8 \pm 0.1) \times \langle \bar q q \rangle \,,
\\
\nonumber && m_s(2 \, \mathrm{GeV}) = 93^{+9}_{-3} \, \mathrm{MeV} \,,
\\
\label{condensates} && \langle \alpha_s^2 G G \rangle = (6.35 \pm 0.35) \times 10^{-2} \, \mathrm{GeV}^4 \,,
\\
\nonumber && M_0^2 = (0.8 \pm 0.2) \, \mathrm{GeV}^2 \,,
\\
\nonumber && \langle g_s \bar q \sigma G q \rangle = -M_0^2 \times \langle \bar q q \rangle \,,
\\
\nonumber && \langle g_s \bar s \sigma G s \rangle = -M_0^2 \times \langle \bar s s \rangle \,.
\end{eqnarray}

\begin{figure}
\includegraphics[width=0.5\textwidth]{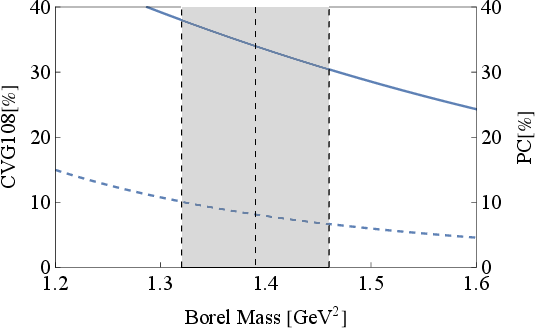}
\caption{Convergence of Dim 10+8 (dashed line) and pole contribution (solid line) under condition of $s_0=4.8$~GeV$^2$, respectively as functions of the Borel mass $M_B^2$.}
\label{fig:figCVGPCMB}
\end{figure}

First, we ensure good convergence of the OPE by requiring that the contribution from dimension-10 and dimension-8 condensates be less than 10\% of the total:
\begin{eqnarray}
\mathrm{CVG} &\equiv& \left| \frac{\Pi^{D=10+8}(\infty, M_B^2)}{\Pi(\infty, M_B^2)} \right| \leq 10\% \,.
\label{eq:convergence}
\end{eqnarray}
This condition imposes a lower bound on the Borel mass: \( M_B^2 \geq 1.32 \, \mathrm{GeV}^2 \).

Second, to ensure the dominance of the ground-state pole contribution, we require that the pole contribution (PC) exceed 30\%:
\begin{equation}
\mathrm{PC} \equiv \left| \frac{\Pi(s_0, M_B^2)}{\Pi(\infty, M_B^2)} \right| \geq 30\% \,.
\label{eq:pole}
\end{equation}
We find that this condition is satisfied only when \( s_0 \geq s_0^{min} = 4.3 \, \mathrm{GeV}^2 \). Accordingly, we adopt a slightly larger value of \( s_0 = 4.8\, \mathrm{GeV}^2 \), resulting in a valid Borel window of \( 1.32 \, \mathrm{GeV}^2 \leq M_B^2 \leq 1.46 \, \mathrm{GeV}^2 \), as shown in Fig.~\ref{fig:figCVGPCMB}.

\begin{figure*}[]
\begin{center}
\subfigure[]{\includegraphics[width=0.45\textwidth]{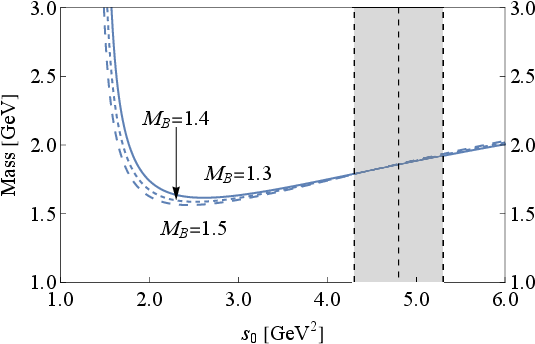}}
~~
\subfigure[]{\includegraphics[width=0.45\textwidth]{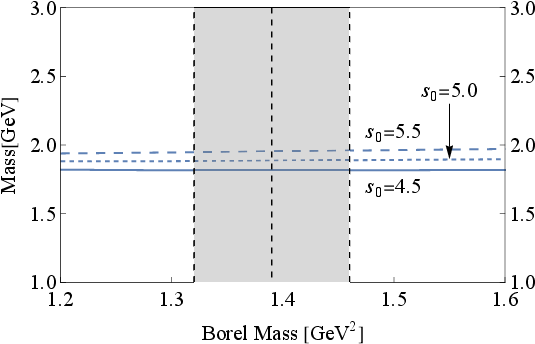}}
\caption{Dependence of the mass on (a) the threshold value \( s_0 \) and (b) the Borel mass \( M_B^2 \). In the left panel, the solid, short-dashed, and long-dashed curves correspond to \( M_B^2 = 1.3,\,1.4,\,1.5~\mathrm{GeV}^2 \), respectively. In the right panel, the curves correspond to \( s_0 = 4.5,\,5.0,\,5.5~\mathrm{GeV}^2 \), respectively.}
\label{fig:figMasss0&figMassMB}
\end{center}
\end{figure*}

Finally, we require that the extracted mass exhibit only mild sensitivity to variations in both the threshold value \( s_0 \) and the Borel mass \( M_B^2 \). As shown in Fig.~\ref{fig:figMasss0&figMassMB}, the mass dependence on these parameters remains moderate within the working regions \( 3.8 \, \mathrm{GeV}^2 \leq s_0 \leq 5.8 \, \mathrm{GeV}^2 \) and \( 1.32 \, \mathrm{GeV}^2 \leq M_B^2 \leq 1.46 \, \mathrm{GeV}^2 \). The resulting mass is determined to be:
\begin{equation}
M =1.86^{+0.14}_{-0.14} \, \mathrm{GeV} \,,
\end{equation}
where the uncertainty reflects the combined variations in the Borel mass \( M_B^2 \), the threshold value \( s_0 \), and the QCD input parameters listed in Eqs.~(\ref{condensates}).

We apply the above procedures to study other currents in a similar manner, and the obtained results are summarized in Table~\ref{tab:result}. QCD sum rule analyses performed on \(q q \bar q \bar q\), \(q s \bar q \bar s\), and \(s s \bar s \bar s\) yield masses of approximately \(1.53 \sim 2.37~\mathrm{GeV}\), \(1.86 \sim 2.51~\mathrm{GeV}\), and \(2.46 \sim 2.60~\mathrm{GeV}\), respectively. Multiple independent currents for each case give similar results, and their mixings also lead to comparable values. Compared to the experimental data, $\psi^M_{2\mu}(q s \bar q \bar s)$ is suitable for the interpretation of the states  with \( I^G J^{PC} = 1^- 1^{--} \). 

Note that the $s s\bar{s}\bar{s}$ tetraquark states with
$I^GJ^{PC}=0^-1^{--}$ have already been systematically investigated
in Refs.~\cite{Chen:2008ej,Chen:2018kuu,Su:2022eun} within the same
QCD sum rule approach. In particular, the lowest mass was obtained in Ref.~\cite{Su:2022eun}
using a current containing one derivative operator,
\begin{equation}
J^{1^{--}}_{7,\alpha}
=
\left[
s_a^T C \gamma_\mu s_b
\right]
\overleftrightarrow{D}_{\alpha}
\left[
\bar{s}_a \gamma^\mu C \bar{s}_b^T
\right] \, ,
\end{equation}
as
\begin{equation}
M^{I=0}_{s s \bar s \bar s} = 2.34^{+0.23}_{-0.30}~{\rm GeV} \, .
\end{equation}
This result is consistent with the masses obtained in the present study.
Since the corresponding current contains one derivative operator and is
expected to better describe the internal structure of the
$s s\bar{s}\bar{s}$ tetraquark state, we shall use this value as the lowest mass in the following discussions.

\begin{table*}[hbtp]
\begin{center}
\renewcommand{\arraystretch}{1.5}
\caption{QCD sum rule results for the light tetraquark states with $J^{PC}=1^{--}$.}
\begin{tabular}{c|c|c|c|c|c|c|c}
\hline\hline
~\multirow{2}{*}{Current}~&~~~~\multirow{2}{*}{State~[$I^{G}J^{PC}$]}~~~~&~~~~~$s_0^{min}$~~~~~&\multicolumn{2}{c|}{Working Regions}&~~~~~Pole~~~~~&~~~~Mass~~~~&~Decay Constant~
\\ \cline{4-5}
& & [${\rm GeV}^2$] & ~$M_B^2~[{\rm GeV}^2]$~ & ~$s_0~[{\rm GeV}^2]$~ & [\%] & [GeV] & [GeV$^5$]
\\ \hline\hline
$\psi^{S(q q \bar q \bar q),\,I=0}_{1\mu}$  & $|q q \bar q \bar q; 0^{+}1^{--}\rangle$ &  $6.8$  &  $1.86$--$2.03$  &  $7.4^{+1.0}_{-1.0}$  &  $30$--$36$  &  $2.37^{+0.12}_{-0.12}$  &  $0.0150^{+0.0047}_{-0.0041}$
\\
$\psi^{S(q q \bar q \bar q),\,I=0}_{2\mu}$  & $|q q \bar q \bar q; 0^{+}1^{--}\rangle$ &  $6.3$  &  $1.77$--$1.92$  &  $6.8^{+1.0}_{-1.0}$  &  $30$--$36$  &  $2.24^{+0.12}_{-0.14}$  &  $0.0152^{+0.0051}_{-0.0044}$
\\
$\psi^{S(q s \bar q \bar s),\,I=0}_{1\mu}$  & $|q s \bar q \bar s; 0^{+}1^{--}\rangle$ &  $7.2$  &  $1.86$--$2.10$  &  $8.0^{+1.0}_{-1.0}$  &  $30$--$39$  &  $2.51^{+0.11}_{-0.11}$  &  $0.0175^{+0.0052}_{-0.0046}$
\\
$\psi^{S(q s \bar q \bar s),\,I=0}_{2\mu}$  & $|q s \bar q \bar s; 0^{+}1^{--}\rangle$ &  $6.7$  &  $1.76$--$1.98$  &  $7.4^{+1.0}_{-1.0}$  &  $30$--$39$  &  $2.39^{+0.11}_{-0.12}$  &  $0.0177^{+0.0056}_{-0.0049}$
\\
$\psi^{S(s s \bar s \bar s),\,I=0}_{1\mu}$  & $|s s \bar s \bar s; 0^{+}1^{--}\rangle$ &  $7.6$  &  $1.83$--$1.86$  &  $7.7^{+1.0}_{-1.0}$  &  $30$--$31$  &  $2.60^{+0.13}_{-0.10}$  &  $0.0163^{+0.0052}_{-0.0044}$
\\
$\psi^{S(s s \bar s \bar s),\,I=0}_{2\mu}$  & $|s s \bar s \bar s; 0^{+}1^{--}\rangle$ &  $7.1$  &  $1.74$--$1.78$  &  $7.2^{+1.0}_{-1.0}$  &  $30$--$32$  &  $2.46^{+0.13}_{-0.11}$  &  $0.0165^{+0.0056}_{-0.0048}$
\\
$\psi^{A(q q \bar q \bar q),\,I=0}_{1\mu}$  & $|q q \bar q \bar q; 0^{+}1^{--}\rangle$ &  $3.3$  &  $1.18$--$1.28$  &  $3.7^{+1.0}_{-1.0}$  &  $30$--$36$  &  $1.64^{+0.15}_{-0.14}$  &  $0.0028^{+0.0013}_{-0.0010}$
\\
$\psi^{A(q q \bar q \bar q),\,I=0}_{2\mu}$  & $|q q \bar q \bar q; 0^{+}1^{--}\rangle$ &  $4.7$  &  $1.47$--$1.60$  &  $5.2^{+1.0}_{-1.0}$  &  $30$--$37$  &  $1.88^{+0.15}_{-0.18}$  &  $0.0119^{+0.0048}_{-0.0041}$
\\
$\psi^{A(q s \bar q \bar s),\,I=0}_{1\mu}$  & $|q s \bar q \bar s; 0^{+}1^{--}\rangle$ &  $4.7$  &  $1.39$--$1.52$  &  $5.2^{+1.0}_{-1.0}$  &  $30$--$37$  &  $1.98^{+0.13}_{-0.12}$  &  $0.0048^{+0.0019}_{-0.0016}$
\\
$\psi^{A(q s \bar q \bar s),\,I=0}_{2\mu}$  & $|q s \bar q \bar s; 0^{+}1^{--}\rangle$ &  $5.2$  &  $1.50$--$1.64$  &  $5.7^{+1.0}_{-1.0}$  &  $30$--$37$  &  $2.03^{+0.14}_{-0.15}$  &  $0.0137^{+0.0054}_{-0.0046}$
\\
$\psi^{M(q s \bar q \bar s),\,I=0}_{1\mu}$  & $|q s \bar q \bar s; 0^{+}1^{--}\rangle$ &  $6.8$  &  $1.77$--$1.99$  &  $7.5^{+1.0}_{-1.0}$  &  $30$--$38$  &  $2.42^{+0.11}_{-0.11}$  &  $0.0150^{+0.0047}_{-0.0041}$
\\
$\psi^{M(q s \bar q \bar s),\,I=0}_{2\mu}$  & $|q s \bar q \bar s; 0^{+}1^{--}\rangle$ &  $4.3$  &  $1.32$--$1.46$  &  $4.8^{+1.0}_{-1.0}$  &  $30$--$38$  &  $1.86^{+0.14}_{-0.14}$  &  $0.0099^{+0.0042}_{-0.0034}$
\\
$\psi^{M(q s \bar q \bar s),\,I=0}_{3\mu}$  & $|q s \bar q \bar s; 0^{+}1^{--}\rangle$ &  $5.9$  &  $1.63$--$1.79$  &  $6.5^{+1.0}_{-1.0}$  &  $30$--$37$  &  $2.21^{+0.12}_{-0.13}$  &  $0.0075^{+0.0027}_{-0.0023}$
\\
$\psi^{M(q s \bar q \bar s),\,I=0}_{4\mu}$  & $|q s \bar q \bar s; 0^{+}1^{--}\rangle$ &  $6.0$  &  $1.62$--$1.80$  &  $6.6^{+1.0}_{-1.0}$  &  $30$--$38$  &  $2.24^{+0.12}_{-0.12}$  &  $0.0136^{+0.0047}_{-0.0040}$
\\ \hline\hline

$\psi^{S(q q \bar q \bar q),\,I=1}_{1\mu}$  & $|q q \bar q \bar q; 1^{-}1^{--}\rangle$ &  $6.8$  &  $1.86$--$2.03$  &  $7.4^{+1.0}_{-1.0}$  &  $30$--$36$  &  $2.37^{+0.12}_{-0.12}$  &  $0.0150^{+0.0047}_{-0.0041}$
\\
$\psi^{S(q q \bar q \bar q),\,I=1}_{2\mu}$  & $|q q \bar q \bar q; 1^{-}1^{--}\rangle$ &  $6.3$  &  $1.77$--$1.92$  &  $6.8^{+1.0}_{-1.0}$  &  $30$--$36$  &  $2.24^{+0.12}_{-0.14}$  &  $0.0152^{+0.0051}_{-0.0044}$
\\
$\psi^{S(q s \bar q \bar s),\,I=1}_{1\mu}$  & $|q s \bar q \bar s; 1^{-}1^{--}\rangle$ &  $7.2$  &  $1.86$--$2.10$  &  $8.0^{+1.0}_{-1.0}$  &  $30$--$39$  &  $2.51^{+0.11}_{-0.11}$  &  $0.0175^{+0.0052}_{-0.0046}$
\\
$\psi^{S(q s \bar q \bar s),\,I=1}_{2\mu}$  & $|q s \bar q \bar s; 1^{-}1^{--}\rangle$ &  $6.7$  &  $1.76$--$1.98$  &  $7.4^{+1.0}_{-1.0}$  &  $30$--$39$  &  $2.39^{+0.11}_{-0.12}$  &  $0.0177^{+0.0056}_{-0.0049}$
\\
$\psi^{A(q s \bar q \bar s),\,I=1}_{1\mu}$  & $|q s \bar q \bar s; 1^{-}1^{--}\rangle$ &  $4.7$  &  $1.39$--$1.52$  &  $5.2^{+1.0}_{-1.0}$  &  $30$--$37$  &  $1.98^{+0.13}_{-0.12}$  &  $0.0048^{+0.0019}_{-0.0016}$
\\
$\psi^{A(q s \bar q \bar s),\,I=1}_{2\mu}$  & $|q s \bar q \bar s; 1^{-}1^{--}\rangle$ &  $5.2$  &  $1.50$--$1.64$  &  $5.7^{+1.0}_{-1.0}$  &  $30$--$37$  &  $2.03^{+0.14}_{-0.15}$  &  $0.0137^{+0.0054}_{-0.0046}$
\\
$\psi^{M(q q \bar q \bar q),\,I=1}_{1\mu}$  & $|q q \bar q \bar q; 1^{-}1^{--}\rangle$ &  $6.1$  &  $1.71$--$1.88$  &  $6.7^{+1.0}_{-1.0}$  &  $30$--$37$  &  $2.24^{+0.12}_{-0.13}$  &  $0.0119^{+0.0040}_{-0.0035}$
\\
$\psi^{M(q q \bar q \bar q),\,I=1}_{2\mu}$  & $|q q \bar q \bar q; 1^{-}1^{--}\rangle$ &  $3.0$  &  $1.13$--$1.26$  &  $3.5^{+1.0}_{-1.0}$  &  $30$--$39$  &  $1.53^{+0.17}_{-0.19}$  &  $0.0060^{+0.0029}_{-0.0024}$
\\
$\psi^{M(q q \bar q \bar q),\,I=1}_{3\mu}$  & $|q q \bar q \bar q; 1^{-}1^{--}\rangle$ &  $5.4$  &  $1.61$--$1.76$  &  $6.0^{+1.0}_{-1.0}$  &  $30$--$37$  &  $2.07^{+0.14}_{-0.15}$  &  $0.0066^{+0.0024}_{-0.0021}$
\\
$\psi^{M(q q \bar q \bar q),\,I=1}_{4\mu}$  & $|q q \bar q \bar q; 1^{-}1^{--}\rangle$ &  $5.2$  &  $1.54$--$1.68$  &  $5.7^{+1.0}_{-1.0}$  &  $30$--$36$  &  $2.03^{+0.13}_{-0.15}$  &  $0.0102^{+0.0039}_{-0.0033}$
\\
$\psi^{M(q s \bar q \bar s),\,I=1}_{1\mu}$  & $|q s \bar q \bar s; 1^{-}1^{--}\rangle$ &  $6.8$  &  $1.77$--$1.99$  &  $7.5^{+1.0}_{-1.0}$  &  $30$--$38$  &  $2.42^{+0.11}_{-0.11}$  &  $0.0150^{+0.0047}_{-0.0041}$
\\
$\psi^{M(q s \bar q \bar s),\,I=1}_{2\mu}$  & $|q s \bar q \bar s; 1^{-}1^{--}\rangle$ &  $4.3$  &  $1.32$--$1.46$  &  $4.8^{+1.0}_{-1.0}$  &  $30$--$38$  &  $1.86^{+0.14}_{-0.14}$  &  $0.0099^{+0.0042}_{-0.0034}$
\\
$\psi^{M(q s \bar q \bar s),\,I=1}_{3\mu}$  & $|q s \bar q \bar s; 1^{-}1^{--}\rangle$ &  $5.9$  &  $1.63$--$1.79$  &  $6.5^{+1.0}_{-1.0}$  &  $30$--$37$  &  $2.21^{+0.12}_{-0.13}$  &  $0.0075^{+0.0027}_{-0.0023}$
\\
$\psi^{M(q s \bar q \bar s),\,I=1}_{4\mu}$  & $|q s \bar q \bar s; 1^{-}1^{--}\rangle$ &  $6.0$  &  $1.62$--$1.80$  &  $6.6^{+1.0}_{-1.0}$  &  $30$--$38$  &  $2.24^{+0.12}_{-0.12}$  &  $0.0136^{+0.0047}_{-0.0040}$
\\ \hline\hline
$\psi^{S(q q \bar q \bar q),\,I=2}_{1\mu}$  & $|q q \bar q \bar q; 2^{+}1^{--}\rangle$ &  $6.8$  &  $1.86$--$2.03$  &  $7.4^{+1.0}_{-1.0}$  &  $30$--$36$  &  $2.37^{+0.12}_{-0.12}$  &  $0.0150^{+0.0047}_{-0.0041}$
\\
$\psi^{S(q q \bar q \bar q),\,I=2}_{2\mu}$  & $|q q \bar q \bar q; 2^{+}1^{--}\rangle$ &  $6.3$  &  $1.77$--$1.92$  &  $6.8^{+1.0}_{-1.0}$  &  $30$--$36$  &  $2.24^{+0.12}_{-0.14}$  &  $0.0152^{+0.0051}_{-0.0044}$
\\ \hline\hline
\end{tabular}
\label{tab:result}
\end{center}
\end{table*}

\section{Summary}
\label{sec:summary}

In this paper we have performed a systematic QCD sum rule study of light tetraquark states with $J^{PC}=1^{--}$ in the diquark--antidiquark picture. We first constructed a complete set of local interpolating currents by classifying their flavor structures into the symmetric $\mathbf{6}_f\otimes\bar{\mathbf{6}}_f$, antisymmetric $\bar{\mathbf{3}}_f\otimes\mathbf{3}_f$, and mixed $(\bar{\mathbf{3}}_f\otimes\bar{\mathbf{6}}_f)\oplus(\mathbf{6}_f\otimes\mathbf{3}_f)$ representations. We then projected these currents onto the isoscalar, isovector, and isotensor channels, and considered the flavor configurations $q q\bar q\bar q$, $q s\bar q\bar s$, and $s s\bar s\bar s$ ($q=u/d$).

We performed QCD sum rule analyses using these tetraquark currents. The resulting masses and decay constants are summarized in Table~\ref{tab:result}. In particular, the lowest masses adopted for the six flavor-isospin sectors are summarized as
\begin{align}
M^{I=0}_{q q \bar q \bar q} &= 1.64^{+0.15}_{-0.14}~{\rm GeV},
\nonumber\\
M^{I=0}_{q s \bar q \bar s} &= 1.86^{+0.14}_{-0.14}~{\rm GeV},
\nonumber\\
M^{I=0}_{s s \bar s \bar s} &= 2.34^{+0.23}_{-0.30}~{\rm GeV},
\nonumber\\
M^{I=1}_{q q \bar q \bar q} &= 1.53^{+0.17}_{-0.19}~{\rm GeV},
\nonumber\\
M^{I=1}_{q s \bar q \bar s} &= 1.86^{+0.14}_{-0.14}~{\rm GeV},
\nonumber\\
M^{I=2}_{q q \bar q \bar q} &= 2.24^{+0.12}_{-0.14}~{\rm GeV}.
\end{align}
Note that the third value, corresponding to the $I=0$ $s s\bar s\bar s$ sector, was obtained in Ref.~\cite{Su:2022eun} using a current containing one derivative operator. This result is consistent with the result obtained in the present study. Since such a derivative current is expected to better describe the internal structure of the $s s\bar s\bar s$ tetraquark state, we use this value as the lowest mass in the following discussions. 

\begin{figure}
\includegraphics[width=0.5\textwidth]{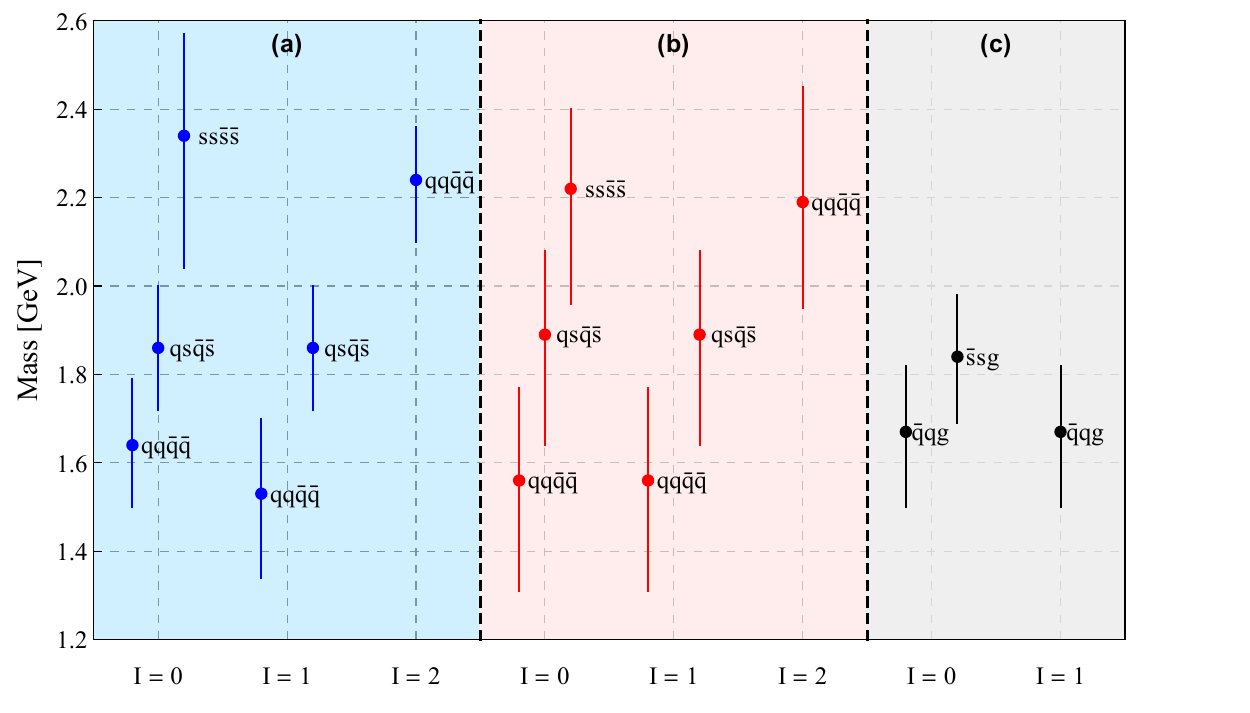}
\caption{Comparison of the mass spectra of light tetraquark and hybrid states. Panel (a) shows the $J^{PC}=1^{--}$ tetraquark masses obtained in the present work, while panels (b) and (c) show our previous results for the $J^{PC}=1^{-+}$ tetraquark and hybrid masses, respectively~\cite{Su:2025bhv}. The tetraquark spectra include six flavor-isospin configurations: the isoscalar $q q\bar q\bar q$, $q s\bar q\bar s$, and $s s\bar s\bar s$ sectors, the isovector $q q\bar q\bar q$ and $q s\bar q\bar s$ sectors, and the isotensor $q q\bar q\bar q$ sector. The hybrid spectrum includes the isoscalar $\bar q q g$ and $\bar s s g$ sectors and the isovector $\bar q q g$ sector.}
\label{fig:figresults}
\end{figure}

The above results are further summarized in Fig.~\ref{fig:figresults}, where we also compare the present $1^{--}$ tetraquark results with our previous results for the $1^{-+}$ tetraquark and hybrid states~\cite{Su:2025bhv}. Both the $1^{--}$ and $1^{-+}$ tetraquark spectra contain six flavor-isospin configurations: the isoscalar $q q\bar q\bar q$, $q s\bar q\bar s$, and $s s\bar s\bar s$ sectors, the isovector $q q\bar q\bar q$ and $q s\bar q\bar s$ sectors, and the isotensor $q q\bar q\bar q$ sector. By contrast, the $1^{-+}$ hybrid spectrum contains three flavor-isospin configurations: the isoscalar $\bar q q g$ and $\bar s s g$ sectors, and the isovector $\bar q q g$ sector. Such a comparison provides a useful reference for understanding how tetraquark and hybrid configurations may populate the light hadron spectrum in different quantum-number channels.

It should be noted that the mass of the isotensor tetraquark state with $J^{PC}=1^{-+}$ was not evaluated in our previous study. In analogy with Sec.~\ref{subsec:isocurrent2}, there are two isotensor $q q\bar q\bar q$ currents with $J^{PC}=1^{-+}$:

\begin{subequations}\label{eq:I2S^*}
\begin{align}
\psi^{S^*(q q \bar q \bar q),\,I=2}_{1\mu}
\equiv{}& \psi^{S^*}_{1\mu}(q q \bar q \bar q)
\label{eq:I2S1s}\\
\sim{}&
u_a^T C \gamma_5 d_b
\left(
\bar{u}_a \gamma_\mu\gamma_5 C \bar{d}_b^T
+\bar{u}_b \gamma_\mu\gamma_5 C \bar{d}_a^T
\right)
\nonumber\\
&+
u_a^T C \gamma_\mu\gamma_5 d_b
\left(
\bar{u}_a \gamma_5 C \bar{d}_b^T
+\bar{u}_b \gamma_5 C \bar{d}_a^T
\right),
\nonumber
\\[0.4em]
\psi^{S^*(q q \bar q \bar q),\,I=2}_{2\mu}
\equiv{}& \psi^{S^*}_{2\mu}(q q \bar q \bar q)
\label{eq:I2S2s}\\
\sim{}&
u_a^T C \gamma^\nu d_b
\left(
\bar{u}_a \sigma_{\mu\nu} C \bar{d}_b^T
-\bar{u}_b \sigma_{\mu\nu} C \bar{d}_a^T
\right)
\nonumber\\
&+
u_a^T C \sigma_{\mu\nu} d_b
\left(
\bar{u}_a \gamma^\nu C \bar{d}_b^T
-\bar{u}_b \gamma^\nu C \bar{d}_a^T
\right).
\nonumber
\end{align}
\end{subequations}
In the present study we apply the same QCD sum rule procedures to these two currents and derive their masses as $2.19^{+0.26}_{-0.24}~\mathrm{GeV}$ and $2.40^{+0.30}_{-0.22}~\mathrm{GeV}$.
The lower of these two mass predictions is included in Fig.~\ref{fig:figresults}, thereby completing the comparison with the $1^{--}$ tetraquark spectrum.

Since $J^{PC}=1^{-+}$ carries exotic quantum numbers, the corresponding hybrid states have attracted considerable interest. However, it remains difficult to distinguish a $1^{-+}$ hybrid state from a $1^{-+}$ tetraquark state, because both configurations may appear in a similar mass region and may couple to similar hadronic channels. Although the $1^{--}$ channel does not carry exotic quantum numbers, it is experimentally important since vector states can be directly produced in electron--positron annihilation. A combined investigation of the $1^{--}$ and $1^{-+}$ channels may therefore provide a broader framework for understanding the similarities and differences between tetraquark and hybrid configurations, and may offer useful clues for distinguishing these two types of exotic hadrons. In the present work we have focused on the $1^{--}$ tetraquark states, while a further study of the corresponding $1^{--}$ hybrid states and the decay properties of these states will be carried out in future work.

\section*{Acknowledgments}
This work is supported by the National Natural Science Foundation of China under Grant No. 12505156 and 12005172.
\appendix

\section{Two-point Correlation Functions}\label{app:ope}
%

In this appendix we present the Borel-transformed correlation functions for all the interpolating currents defined in Sec.~\ref{sec:current}.

\begin{widetext}
\begin{eqnarray}\nonumber
\Pi^{S(q q \bar q \bar q)}_{1\mu}(M_B^2) &=& \int^{s_0}_{0} \Bigg [ 
{ 1 \over 18432 \pi^6 } s^4
- { \langle g_s^2 G G \rangle \over 18432 \pi^6 } s^2 
+ { \langle \bar q q \rangle^2 \over 18 \pi^2 } s
+ { \langle \bar q q \rangle \langle g_s \bar q \sigma G q \rangle \over 8 \pi^2 }
\Bigg ] e^{-s/M_B^2} ds
\\ \nonumber && + \Big (
{ 5 \langle g_s^2 G G \rangle \langle \bar q q \rangle ^2 \over 864 \pi^2 }
+ { \langle g_s \bar q \sigma G q \rangle^2 \over 24 \pi^2}
\Big ),
\end{eqnarray}
\begin{eqnarray}\nonumber
\Pi^{S(q q \bar q \bar q)}_{2\mu}(M_B^2) &=& \int^{s_0}_{0} \Bigg [
{ 1 \over 12288 \pi^6 } s^4
+ { \langle g_s^2 G G \rangle \over 18432 \pi^6 } s^2
+ { \langle \bar q q \rangle^2 \over 12 \pi^2 } s
+ { \langle \bar q q \rangle \langle g_s \bar q \sigma G q \rangle \over 6 \pi^2 }
\Bigg ] e^{-s/M_B^2} ds
\\ \nonumber && + \Big (
{ 5 \langle g_s^2 G G \rangle \langle \bar q q \rangle ^2 \over 288 \pi^2 }
+ { 5 \langle g_s \bar q \sigma G q \rangle^2 \over 96 \pi^2 }
\Big ),
\end{eqnarray}
\begin{eqnarray}\nonumber
\Pi^{S(q s \bar q \bar s)}_{1\mu}(M_B^2) &=& \int^{s_0}_{4m_s^2} \Bigg [
{ 1 \over 18432 \pi^6 } s^4
- { m_s^2 \over 512 \pi^6 } s^3
+ \Big (
- { \langle g_s^2 G G \rangle \over 18432 \pi^6 }
+ { 3 m_s^4 \over 256 \pi^6 } 
+ { m_s \langle \bar s s \rangle \over 96 \pi^4 }
\Big ) s^2 
\\ \nonumber && + \Big (
{ m_s^2 \langle g_s^2 G G \rangle \over 4608 \pi^6 }
- { m_s^3 \langle \bar s s \rangle \over 12 \pi^4 }
- { m_s \langle g_s \bar q \sigma G q \rangle \over 384 \pi^4 }
- { 5 m_s \langle g_s \bar s \sigma G s \rangle \over 384 \pi^4 }
+ { \langle \bar q q \rangle^2 \over 36 \pi^2 }
+ { \langle \bar s s \rangle^2 \over 36 \pi^2 }
\Big ) s
\\ \nonumber && + \Big (
- { m_s \langle g_s^2 G G \rangle \langle \bar q q \rangle \over 256 \pi^4 }
- { m_s^2 \langle \bar q q \rangle^2 \over 2 \pi^2 }
+ { m_s^2 \langle \bar s s \rangle^2 \over 24 \pi^2 }
+ { 5 \langle \bar q q \rangle \langle g_s \bar q \sigma G q \rangle \over 96 \pi^2 } 
+ { \langle \bar q q \rangle \langle g_s \bar s \sigma G s \rangle \over 96 \pi^2 } 
+ { \langle \bar s s \rangle \langle g_s \bar q \sigma G q \rangle \over 96 \pi^2 }
\\ \nonumber && + { 5 \langle \bar s s \rangle \langle g_s \bar s \sigma G s \rangle \over 96 \pi^2 }
\Big )
\Bigg ] e^{-s/M_B^2} ds
+ \Big (
- { 5 m_s \langle g_s^2 G G \rangle \langle g_s \bar q \sigma G q \rangle \over 4608 \pi^4 }
+ { 5 \langle g_s^2 G G \rangle \langle \bar q q \rangle \langle \bar s s \rangle \over 864 \pi^2 }
+ { m_s^4 \langle \bar q q \rangle^2 \over 6 \pi^2 }
\\ \nonumber && - { 17 m_s^2 \langle \bar q q \rangle \langle g_s \bar q \sigma G q \rangle \over 48 \pi^2 }
- { m_s^2 \langle \bar q q \rangle \langle g_s \bar s \sigma G s \rangle \over 48 \pi^2 }
+ { 10 m_s \langle \bar q q \rangle^2 \langle \bar s s \rangle \over 9 }
+ { \langle g_s \bar q \sigma G q \rangle^2 \over 64 \pi^2 }
+ { \langle g_s \bar q \sigma G q \rangle \langle g_s \bar s \sigma G s \rangle \over 96 \pi^2 }
\\ \nonumber && + { \langle g_s \bar s \sigma G s \rangle^2 \over 64 \pi^2 }
\Big ),
\end{eqnarray}
\begin{eqnarray}\nonumber
\Pi^{S(q s \bar q \bar s)}_{2\mu}(M_B^2) &=& \int^{s_0}_{4m_s^2} \Bigg [
{ 1 \over 12288 \pi^6 } s^4
- { 3 m_s^2 \over 1024 \pi^6 } s^3
+ \Big (
{ \langle g_s^2 G G \rangle \over 18432 \pi^6 }
+ { 9 m_s^4 \over 512 \pi^6 } 
+ { m_s \langle \bar s s \rangle \over 64 \pi^4 }
\Big ) s^2 
\\ \nonumber && + \Big (
{ m_s^2 \langle g_s^2 G G \rangle \over 2048 \pi^6 }
- { m_s^3 \langle \bar s s \rangle \over 8 \pi^4 }
- { m_s \langle g_s \bar q \sigma G q \rangle \over 192 \pi^4 }
- { m_s \langle g_s \bar s \sigma G s \rangle \over 64 \pi^4 }
+ { \langle \bar q q \rangle^2 \over 24 \pi^2 }
+ { \langle \bar s s \rangle^2 \over 24 \pi^2 }
\Big ) s
\\ \nonumber && + \Big (
- { m_s \langle g_s^2 G G \rangle \langle \bar q q \rangle \over 128 \pi^4 }
- { m_s \langle g_s^2 G G \rangle \langle \bar s s \rangle \over 256 \pi^4 }
- { 3 m_s^2 \langle \bar q q \rangle^2 \over 4 \pi^2 }
+ { m_s^2 \langle \bar s s \rangle^2 \over 16 \pi^2 }
+ { \langle \bar q q \rangle \langle g_s \bar q \sigma G q \rangle \over 16 \pi^2 }
+ { \langle \bar q q \rangle \langle g_s \bar s \sigma G s \rangle \over 48 \pi^2 }
\\ \nonumber && + { \langle \bar s s \rangle \langle g_s \bar q \sigma G q \rangle \over 48 \pi^2 }
+ { \langle \bar s s \rangle \langle g_s \bar s \sigma G s \rangle \over 16 \pi^2 }
\Big )
\Bigg ] e^{-s/M_B^2} ds
+ \Big (
{ m_s^3 \langle g_s^2 G G \rangle \langle \bar s s \rangle \over 1152 \pi^2 }
- { 5 m_s \langle g_s^2 G G \rangle \langle g_s \bar q \sigma G q \rangle \over 2304 \pi^4 }
\\ \nonumber && - { 5 m_s \langle g_s^2 G G \rangle \langle g_s \bar s \sigma G s \rangle \over 4608 \pi^4 }
+ { 5 \langle g_s^2 G G \rangle \langle \bar q q \rangle^2 \over 1728 \pi^2 }
+ { 5 \langle g_s^2 G G \rangle \langle \bar q q \rangle \langle \bar s s \rangle \over 432 \pi^2 }
+ { 5 \langle g_s^2 G G \rangle \langle \bar s s \rangle^2 \over 1728 \pi^2 }
+ { m_s^4 \langle \bar q q \rangle^2 \over 4 \pi^2 }
\\ \nonumber && - { m_s^2 \langle \bar q q \rangle \langle g_s \bar q \sigma G q \rangle \over 2 \pi^2 }
- { m_s^2 \langle \bar q q \rangle \langle g_s \bar s \sigma G s \rangle \over 24 \pi^2 }
+ { 5 m_s \langle \bar q q \rangle^2 \langle \bar s s \rangle \over 3 }
+ { \langle g_s \bar q \sigma G q \rangle^2 \over 64 \pi^2 }
+ { \langle g_s \bar q \sigma G q \rangle \langle g_s \bar s \sigma G s \rangle \over 48 \pi^2 }
\\ \nonumber && + { \langle g_s \bar s \sigma G s \rangle^2 \over 64 \pi^2 }
\Big ),
\end{eqnarray}
\begin{eqnarray}\nonumber
\Pi^{S(s s \bar s \bar s)}_{1\mu}(M_B^2) &=& \int^{s_0}_{16m_s^2} \Bigg [
{ 1 \over 18432 \pi^6 } s^4
- { m_s^2 \over 256 \pi^6 } s^3
+ \Big (
- { \langle g_s^2 G G \rangle \over 18432 \pi^6 }
+ { 3 m_s^4 \over 64 \pi^6 } 
+ { m_s \langle \bar s s \rangle \over 48 \pi^4 }
\Big ) s^2 
\\ \nonumber && + \Big (
{ 17 m_s^2 \langle g_s^2 GG \rangle \over 9216 \pi^6 }
- { 7 m_s^6 \over 32 \pi^6 }
+ { 5 m_s^3 \langle \bar s s \rangle \over 24 \pi^4 }
- { m_s \langle g_s \bar s \sigma G s \rangle \over 32 \pi^4 }
+ { \langle \bar s s \rangle^2 \over 18 \pi^2 }
\Big ) s
\\ \nonumber && + \Big (
{ 3 m_s^8 \over 32 \pi^6 }
- { m_s \langle g_s^2 G G \rangle \langle \bar s s \rangle \over 128 \pi^4 }
+ { 3 m_s^5 \langle \bar s s \rangle \over 8 \pi^4 }
- { 29 m_s^2 \langle \bar s s \rangle^2 \over 12 \pi^2 }
+ { 13 m_s^3 \langle g_s \bar s \sigma G s \rangle \over 32 \pi^4 }
+ { \langle \bar s s \rangle \langle g_s \bar s \sigma G s \rangle \over 8 \pi^2 }
\Big )
\Bigg ] e^{-s/M_B^2} ds
\\ \nonumber && + \Big (
- { m_s^3 \langle g_s^2 G G \rangle \langle \bar s s \rangle \over 576 \pi^4 }
- { 5 m_s \langle g_s^2 G G \rangle \langle g_s \bar s \sigma G s \rangle \over 2304 \pi^4 }
+ { 5 \langle g_s^2 G G \rangle \langle \bar s s \rangle^2 \over 864 \pi^2 }
- { m_s^5 \langle g_s \bar s \sigma G s \rangle \over 8 \pi^4 }
+ { 13 m_s^4 \langle \bar s s \rangle^2 \over 12 \pi^2 }
\\ \nonumber && - { 13 m_s^2 \langle \bar s s \rangle \langle g_s \bar s \sigma G s \rangle \over 8 \pi^2 }
+ { 20 m_s \langle \bar s s \rangle^3 \over 9 }
+ { \langle g_s \bar s \sigma G s \rangle^2 \over 24 \pi^2 }
\Big ),
\end{eqnarray}
\begin{eqnarray}\nonumber
\Pi^{S(s s \bar s \bar s)}_{2\mu}(M_B^2) &=& \int^{s_0}_{16m_s^2} \Bigg [
{ 1 \over 12288 \pi^6 } s^4
- { 3 m_s^2 \over 512 \pi^6 } s^3
+ \Big (
{ \langle g_s^2 G G \rangle \over 18432 \pi^6 }
+ { 9 m_s^4 \over 128 \pi^6 } 
+ { m_s \langle \bar s s \rangle \over 32 \pi^4 }
\Big ) s^2 
\\ \nonumber && + \Big (
{ 35 m_s^2 \langle g_s^2 G G \rangle \over 9216 \pi^6 }
- { 21 m_s^6 \over 64 \pi^6 }
+ { 5 m_s^3 \langle \bar s s \rangle \over 16 \pi^4 }
- { m_s \langle g_s \bar s \sigma G s \rangle \over 24 \pi^4 }
+ { \langle \bar s s \rangle^2 \over 12 \pi^2 }
\Big ) s
\\ \nonumber && + \Big (
- { 3 m_s \langle g_s^2 G G \rangle \langle \bar s s \rangle \over 128 \pi^4 }
+ { 9 m_s^8 \over 64 \pi^6 }
+ { 9 m_s^5 \langle \bar s s \rangle \over 16 \pi^4 }
+ { 19 m_s^3 \langle g_s \bar s \sigma G s \rangle \over 32 \pi^4 }
- { 29 m_s^2 \langle \bar s s \rangle^2 \over 8 \pi^2 }
\\ \nonumber && + { \langle \bar s s \rangle \langle g_s \bar s \sigma G s \rangle \over 6 \pi^2 }
\Big )
\Bigg ] e^{-s/M_B^2} ds
+ \Big (
{ m_s^3 \langle g_s^2 G G \rangle \langle \bar s s \rangle \over 576 \pi^4 }
- { 5 m_s \langle g_s^2 G G \rangle \langle g_s \bar s \sigma G s \rangle \over 768 \pi^4 }
+ { 5 \langle g_s^2 G G \rangle \langle \bar s s \rangle^2 \over 288 \pi^2 }
\\ \nonumber && - { 3 m_s^5 \langle g_s \bar s \sigma G s \rangle \over 16 \pi^4 }
+ { 13 m_s^4 \langle \bar s s \rangle^2 \over 8 \pi^2 }
- { 19 m_s^2 \langle \bar s s \rangle \langle g_s \bar s \sigma G s \rangle \over 8 \pi^2 }
+ { 10 m_s \langle \bar s s \rangle^3 \over 3 }
+ { 5 \langle g_s \bar s \sigma G s \rangle^2 \over 96 \pi^2 }
\Big ),
\end{eqnarray}
\begin{eqnarray}\nonumber
\Pi^{A(q q \bar q \bar q)}_{1\mu}(M_B^2) &=& \int^{s_0}_{0} \Bigg [ 
{ 1 \over 36864 \pi^6 } s^4
+ { \langle g_s^2 G G \rangle \over 18432 \pi^6 } s^2 
+ { \langle \bar q q \rangle^2 \over 36 \pi^2 } s
\Bigg ] e^{-s/M_B^2} ds
\\ \nonumber && + \Big (
- { 5 \langle g_s^2 G G \rangle \langle \bar q q \rangle ^2 \over 864 \pi^2 }
- { \langle g_s \bar q \sigma G q \rangle^2 \over 96 \pi^2}
\Big ),
\end{eqnarray}
\begin{eqnarray}\nonumber
\Pi^{A(q q \bar q \bar q)}_{2\mu}(M_B^2) &=& \int^{s_0}_{0} \Bigg [
{ 1 \over 6144 \pi^6 } s^4
+ { 11 \langle g_s^2 G G \rangle \over 18432 \pi^6 } s^2
+ { \langle \bar q q \rangle^2 \over 6 \pi^2 } s
+ { 5 \langle \bar q q \rangle \langle g_s \bar q \sigma G q \rangle \over 24 \pi^2 }
\Bigg ] e^{-s/M_B^2} ds
\\ \nonumber && + \Big (
{ 5 \langle g_s^2 G G \rangle \langle \bar q q \rangle ^2 \over 96 \pi^2 }
+ { \langle g_s \bar q \sigma G q \rangle^2 \over 24 \pi^2 }
\Big ),
\end{eqnarray}
\begin{eqnarray}\nonumber
\Pi^{A(q s \bar q \bar s)}_{1\mu}(M_B^2) &=& \int^{s_0}_{4m_s^2} \Bigg [
{ 1 \over 36864 \pi^6 } s^4
- { m_s^2 \over 1024 \pi^6 } s^3
+ \Big (
{ \langle g_s^2 G G \rangle \over 18432 \pi^6 }
+ { 3 m_s^4 \over 512 \pi^6 } 
+ { m_s \langle \bar s s \rangle \over 192 \pi^4 }
\Big ) s^2 
\\ \nonumber && + \Big (
- { m_s^2 \langle g_s^2 G G \rangle \over 4608 \pi^6 }
- { m_s^3 \langle \bar s s \rangle \over 24 \pi^4 }
+ { m_s \langle g_s \bar q \sigma G q \rangle \over 384 \pi^4 }
- { m_s \langle g_s \bar s \sigma G s \rangle \over 384 \pi^4 }
+ { \langle \bar q q \rangle^2 \over 72 \pi^2 }
+ { \langle \bar s s \rangle^2 \over 72 \pi^2 }
\Big ) s
\\ \nonumber && + \Big (
{ m_s \langle g_s^2 G G \rangle \langle \bar q q \rangle \over 256 \pi^4 }
- { m_s^2 \langle \bar q q \rangle^2 \over 4 \pi^2 }
+ { m_s^2 \langle \bar s s \rangle^2 \over 48 \pi^2 }
+ { \langle \bar q q \rangle \langle g_s \bar q \sigma G q \rangle \over 96 \pi^2 } 
- { \langle \bar q q \rangle \langle g_s \bar s \sigma G s \rangle \over 96 \pi^2 } 
- { \langle \bar s s \rangle \langle g_s \bar q \sigma G q \rangle \over 96 \pi^2 } 
\\ \nonumber && + { \langle \bar s s \rangle \langle g_s \bar s \sigma G s \rangle \over 96 \pi^2 }
\Big )
\Bigg ] e^{-s/M_B^2} ds
+ \Big (
{ 5 m_s \langle g_s^2 G G \rangle \langle g_s \bar q \sigma G q \rangle \over 4608 \pi^4 }
- { 5 \langle g_s^2 G G \rangle \langle \bar q q \rangle \langle \bar s s \rangle \over 864 \pi^2 }
+ { m_s^4 \langle \bar q q \rangle^2 \over 12 \pi^2 }
\\ \nonumber && - { 7 m_s^2 \langle \bar q q \rangle \langle g_s \bar q \sigma G q \rangle \over 48 \pi^2 }
+ { m_s^2 \langle \bar q q \rangle \langle g_s \bar s \sigma G s \rangle \over 48 \pi^2 }
+ { 5 m_s \langle \bar q q \rangle^2 \langle \bar s s \rangle \over 9 }
- { \langle g_s \bar q \sigma G q \rangle \langle g_s \bar s \sigma G s \rangle \over 96 \pi^2 }
\Big ),
\end{eqnarray}
\begin{eqnarray}\nonumber
\Pi^{A(q s \bar q \bar s)}_{2\mu}(M_B^2) &=& \int^{s_0}_{4m_s^2} \Bigg [
{ 1 \over 6144 \pi^6 } s^4
- { 3 m_s^2 \over 512 \pi^6 } s^3
+ \Big (
{ 11 \langle g_s^2 G G \rangle \over 18432 \pi^6 }
+ { 9 m_s^4 \over 256 \pi^6 } 
+ { m_s \langle \bar s s \rangle \over 32 \pi^4 }
\Big ) s^2 
\\ \nonumber && + \Big (
{ 7 m_s^2 \langle g_s^2 G G \rangle \over 6144 \pi^6 }
- { m_s^3 \langle \bar s s \rangle \over 4 \pi^4 }
- { m_s \langle g_s \bar q \sigma G q \rangle \over 96 \pi^4 }
- { m_s \langle g_s \bar s \sigma G s \rangle \over 64 \pi^4 }
+ { \langle \bar q q \rangle^2 \over 12 \pi^2 }
+ { \langle \bar s s \rangle^2 \over 12 \pi^2 }
\Big ) s
\\ \nonumber && + \Big (
- { m_s \langle g_s^2 G G \rangle \langle \bar q q \rangle \over 64 \pi^4 }
- { 5 m_s \langle g_s^2 G G \rangle \langle \bar s s \rangle \over 256 \pi^4 }
- { 3 m_s^2 \langle \bar q q \rangle^2 \over 2 \pi^2 }
+ { m_s^2 \langle \bar s s \rangle^2 \over 8 \pi^2 }
+ { \langle \bar q q \rangle \langle g_s \bar q \sigma G q \rangle \over 16 \pi^2 }
+ { \langle \bar q q \rangle \langle g_s \bar s \sigma G s \rangle \over 24 \pi^2 }
\\ \nonumber && + { \langle \bar s s \rangle \langle g_s \bar q \sigma G q \rangle \over 24 \pi^2 }
+ { \langle \bar s s \rangle \langle g_s \bar s \sigma G s \rangle \over 16 \pi^2 }
\Big )
\Bigg ] e^{-s/M_B^2} ds
+ \Big (
{ 5 m_s^3 \langle g_s^2 G G \rangle \langle \bar s s \rangle \over 1152 \pi^2 }
- { 5 m_s \langle g_s^2 G G \rangle \langle g_s \bar q \sigma G q \rangle \over 1152 \pi^4 }
\\ \nonumber && - { 25 m_s \langle g_s^2 G G \rangle \langle g_s \bar s \sigma G s \rangle \over 4608 \pi^4 }
+ { 25 \langle g_s^2 G G \rangle \langle \bar q q \rangle^2 \over 1728 \pi^2 }
+ { 5 \langle g_s^2 G G \rangle \langle \bar q q \rangle \langle \bar s s \rangle \over 216 \pi^2 }
+ { 25 \langle g_s^2 G G \rangle \langle \bar s s \rangle^2 \over 1728 \pi^2 }
+ { m_s^4 \langle \bar q q \rangle^2 \over 2 \pi^2 }
\\ \nonumber && - { 7 m_s^2 \langle \bar q q \rangle \langle g_s \bar q \sigma G q \rangle \over 8 \pi^2 }
- { m_s^2 \langle \bar q q \rangle \langle g_s \bar q \sigma G q \rangle \over 12 \pi^2 }
+ { 10 m_s \langle \bar q q \rangle^2 \langle \bar s s \rangle \over 3 }
+ { \langle g_s \bar q \sigma G q \rangle \langle g_s \bar s \sigma G s \rangle \over 24 \pi^2 }
\Big ),
\end{eqnarray}
\begin{eqnarray}\nonumber
\Pi^{M(q q \bar q \bar q)}_{1\mu}(M_B^2) &=& \int^{s_0}_{0} \Bigg [ 
{ 1 \over 18432 \pi^6 } s^4
- { \langle g_s^2 G G \rangle \over 18432 \pi^6 } s^2 
+ { \langle \bar q q \rangle^2 \over 18 \pi^2 } s
+ { \langle \bar q q \rangle \langle g_s \bar q \sigma G q \rangle \over 12 \pi^2 }
\Bigg ] e^{-s/M_B^2} ds
\\ \nonumber && + \Big (
- { 5 \langle g_s^2 G G \rangle \langle \bar q q \rangle ^2 \over 864 \pi^2 }
+ { \langle g_s \bar q \sigma G q \rangle^2 \over 48 \pi^2}
\Big ),
\end{eqnarray}
\begin{eqnarray}\nonumber
\Pi^{M(q q \bar q \bar q)}_{2\mu}(M_B^2) &=& \int^{s_0}_{0} \Bigg [
{ 1 \over 6144 \pi^6 } s^4
+ { 11 \langle g_s^2 G G \rangle \over 18432 \pi^6 } s^2
+ { \langle \bar q q \rangle^2 \over 6 \pi^2 } s
+ { \langle \bar q q \rangle \langle g_s \bar q \sigma G q \rangle \over 24 \pi^2 }
\Bigg ] e^{-s/M_B^2} ds
\\ \nonumber && + \Big (
{ 5 \langle g_s^2 G G \rangle \langle \bar q q \rangle ^2 \over 864 \pi^2 }
- { \langle g_s \bar q \sigma G q \rangle^2 \over 24 \pi^2 }
\Big ),
\end{eqnarray}
\begin{eqnarray}\nonumber
\Pi^{M(q q \bar q \bar q)}_{3\mu}(M_B^2) &=& \int^{s_0}_{0} \Bigg [ 
{ 1 \over 36864 \pi^6 } s^4
+ { \langle g_s^2 G G \rangle \over 18432 \pi^6 } s^2 
+ { \langle \bar q q \rangle^2 \over 36 \pi^2 } s
+ { \langle \bar q q \rangle \langle g_s \bar q \sigma G q \rangle \over 24 \pi^2 }
\Bigg ] e^{-s/M_B^2} ds
\\ \nonumber && + \Big (
{ 5 \langle g_s^2 G G \rangle \langle \bar q q \rangle ^2 \over 864 \pi^2 }
+ { \langle g_s \bar q \sigma G q \rangle^2 \over 96 \pi^2}
\Big ),
\end{eqnarray}
\begin{eqnarray}\nonumber
\Pi^{M(q q \bar q \bar q)}_{4\mu}(M_B^2) &=& \int^{s_0}_{0} \Bigg [
{ 1 \over 12288 \pi^6 } s^4
+ { \langle g_s^2 G G \rangle \over 18432 \pi^6 } s^2
+ { \langle \bar q q \rangle^2 \over 12 \pi^2 } s
+ { \langle \bar q q \rangle \langle g_s \bar q \sigma G q \rangle \over 12 \pi^2 }
\Bigg ] e^{-s/M_B^2} ds
\\ \nonumber && + \Big (
- { 5 \langle g_s^2 G G \rangle \langle \bar q q \rangle ^2 \over 864 \pi^2 }
+ { \langle g_s \bar q \sigma G q \rangle^2 \over 96 \pi^2 }
\Big ),
\end{eqnarray}
\begin{eqnarray}\nonumber
\Pi^{M(q s \bar q \bar s)}_{1\mu}(M_B^2) &=& \int^{s_0}_{4m_s^2} \Bigg [
{ 1 \over 18432 \pi^6 } s^4
- { m_s^2 \over 512 \pi^6 } s^3
+ \Big (
- { \langle g_s^2 G G \rangle \over 18432 \pi^6 }
+ { 3 m_s^4 \over 256 \pi^6 } 
+ { m_s \langle \bar s s \rangle \over 96 \pi^4 }
\Big ) s^2 
\\ \nonumber && + \Big (
{ m_s^2 \langle g_s^2 G G \rangle \over 4608 \pi^6 }
- { m_s^3 \langle \bar s s \rangle \over 12 \pi^4 }
+ { m_s \langle g_s \bar q \sigma G q \rangle \over 384 \pi^4 }
- { 5 m_s \langle g_s \bar s \sigma G s \rangle \over 384 \pi^4 }
+ { \langle \bar q q \rangle^2 \over 36 \pi^2 }
+ { \langle \bar s s \rangle^2 \over 36 \pi^2 }
\Big ) s
\\ \nonumber && + \Big (
{ m_s \langle g_s^2 G G \rangle \langle \bar q q \rangle \over 256 \pi^4 }
- { m_s^2 \langle \bar q q \rangle^2 \over 2 \pi^2 }
+ { m_s^2 \langle \bar s s \rangle^2 \over 24 \pi^2 }
+ { 5 \langle \bar q q \rangle \langle g_s \bar q \sigma G q \rangle \over 96 \pi^2 } 
- { \langle \bar q q \rangle \langle g_s \bar s \sigma G s \rangle \over 96 \pi^2 } 
- { \langle \bar s s \rangle \langle g_s \bar q \sigma G q \rangle \over 96 \pi^2 } 
\\ \nonumber && + { 5 \langle \bar s s \rangle \langle g_s \bar s \sigma G s \rangle \over 96 \pi^2 }
\Big )
\Bigg ] e^{-s/M_B^2} ds
+ \Big (
{ 5 m_s \langle g_s^2 G G \rangle \langle g_s \bar q \sigma G q \rangle \over 4608 \pi^4 }
- { 5 \langle g_s^2 G G \rangle \langle \bar q q \rangle \langle \bar s s \rangle \over 864 \pi^2 }
+ { m_s^4 \langle \bar q q \rangle^2 \over 6 \pi^2 }
\\ \nonumber && - { 17 m_s^2 \langle \bar q q \rangle \langle g_s \bar q \sigma G q \rangle \over 48 \pi^2 }
+ { m_s^2 \langle \bar q q \rangle \langle g_s \bar s \sigma G s \rangle \over 48 \pi^2 }
+ { 10 m_s \langle \bar q q \rangle^2 \langle \bar s s \rangle \over 9 }
+ { \langle g_s \bar q \sigma G q \rangle^2 \over 64 \pi^2 }
- { \langle g_s \bar q \sigma G q \rangle \langle g_s \bar s \sigma G s \rangle \over 96 \pi^2 }
\\ \nonumber && + { \langle g_s \bar s \sigma G s \rangle^2 \over 64 \pi^2 }
\Big ),
\end{eqnarray}
\begin{eqnarray}\nonumber
\Pi^{M(q s \bar q \bar s)}_{2\mu}(M_B^2) &=& \int^{s_0}_{4m_s^2} \Bigg [
{ 1 \over 6144 \pi^6 } s^4
- { 3 m_s^2 \over 512 \pi^6 } s^3
+ \Big (
{ 11 \langle g_s^2 G G \rangle \over 18432 \pi^6 }
+ { 9 m_s^4 \over 256 \pi^6 } 
+ { m_s \langle \bar s s \rangle \over 32 \pi^4 }
\Big ) s^2 
\\ \nonumber && + \Big (
{ 7 m_s^2 \langle g_s^2 G G \rangle \over 6144 \pi^6 }
- { m_s^3 \langle \bar s s \rangle \over 4 \pi^4 }
+ { m_s \langle g_s \bar q \sigma G q \rangle \over 96 \pi^4 }
- { m_s \langle g_s \bar s \sigma G s \rangle \over 64 \pi^4 }
+ { \langle \bar q q \rangle^2 \over 12 \pi^2 }
+ { \langle \bar s s \rangle^2 \over 12 \pi^2 }
\Big ) s
\\ \nonumber && + \Big (
{ m_s \langle g_s^2 G G \rangle \langle \bar q q \rangle \over 64 \pi^4 }
- { 5 m_s \langle g_s^2 G G \rangle \langle \bar s s \rangle \over 256 \pi^4 }
- { 3 m_s^2 \langle \bar q q \rangle^2 \over 2 \pi^2 }
+ { m_s^2 \langle \bar s s \rangle^2 \over 8 \pi^2 }
+ { \langle \bar q q \rangle \langle g_s \bar q \sigma G q \rangle \over 16 \pi^2 }
- { \langle \bar q q \rangle \langle g_s \bar s \sigma G s \rangle \over 24 \pi^2 }
\\ \nonumber && - { \langle \bar s s \rangle \langle g_s \bar q \sigma G q \rangle \over 24 \pi^2 }
+ { \langle \bar s s \rangle \langle g_s \bar s \sigma G s \rangle \over 16 \pi^2 }
\Big )
\Bigg ] e^{-s/M_B^2} ds
+ \Big (
{ 5 m_s^3 \langle g_s^2 G G \rangle \langle \bar s s \rangle \over 1152 \pi^2 }
+ { 5 m_s \langle g_s^2 G G \rangle \langle g_s \bar q \sigma G q \rangle \over 1152 \pi^4 }
\\ \nonumber && - { 25 m_s \langle g_s^2 G G \rangle \langle g_s \bar s \sigma G s \rangle \over 4608 \pi^4 }
+ { 25 \langle g_s^2 G G \rangle \langle \bar q q \rangle^2 \over 1728 \pi^2 }
- { 5 \langle g_s^2 G G \rangle \langle \bar q q \rangle \langle \bar s s \rangle \over 216 \pi^2 }
+ { 25 \langle g_s^2 G G \rangle \langle \bar s s \rangle^2 \over 1728 \pi^2 }
+ { m_s^4 \langle \bar q q \rangle^2 \over 2 \pi^2 }
\\ \nonumber && - { 7 m_s^2 \langle \bar q q \rangle \langle g_s \bar q \sigma G q \rangle \over 8 \pi^2 }
+ { m_s^2 \langle \bar q q \rangle \langle g_s \bar s \sigma G s \rangle \over 12 \pi^2 }
+ { 10 m_s \langle \bar q q \rangle^2 \langle \bar s s \rangle \over 3 }- { \langle g_s \bar q \sigma G q \rangle \langle g_s \bar s \sigma G s \rangle \over 24 \pi^2 }
\Big ),
\end{eqnarray}
\begin{eqnarray}\nonumber
\Pi^{M(q s \bar q \bar s)}_{3\mu}(M_B^2) &=& \int^{s_0}_{4m_s^2} \Bigg [
{ 1 \over 36864 \pi^6 } s^4
- { m_s^2 \over 1024 \pi^6 } s^3
+ \Big (
{ \langle g_s^2 G G \rangle \over 18432 \pi^6 }
+ { 3 m_s^4 \over 512 \pi^6 } 
+ { m_s \langle \bar s s \rangle \over 192 \pi^4 }
\Big ) s^2 
\\ \nonumber && + \Big (
- { m_s^2 \langle g_s^2 G G \rangle \over 4608 \pi^6 }
- { m_s^3 \langle \bar s s \rangle \over 24 \pi^4 }
- { m_s \langle g_s \bar q \sigma G q \rangle \over 384 \pi^4 }
- { m_s \langle g_s \bar s \sigma G s \rangle \over 384 \pi^4 }
+ { \langle \bar q q \rangle^2 \over 72 \pi^2 }
+ { \langle \bar s s \rangle^2 \over 72 \pi^2 }
\Big ) s
\\ \nonumber && + \Big (
- { m_s \langle g_s^2 G G \rangle \langle \bar q q \rangle \over 256 \pi^4 }
- { m_s^2 \langle \bar q q \rangle^2 \over 4 \pi^2 }
+ { m_s^2 \langle \bar s s \rangle^2 \over 48 \pi^2 }
+ { \langle \bar q q \rangle \langle g_s \bar q \sigma G q \rangle \over 96 \pi^2 } 
+ { \langle \bar q q \rangle \langle g_s \bar s \sigma G s \rangle \over 96 \pi^2 } 
+ { \langle \bar s s \rangle \langle g_s \bar q \sigma G q \rangle \over 96 \pi^2 } 
\\ \nonumber && + { \langle \bar s s \rangle \langle g_s \bar s \sigma G s \rangle \over 96 \pi^2 }
\Big )
\Bigg ] e^{-s/M_B^2} ds
\Big (
- { 5 m_s \langle g_s^2 G G \rangle \langle g_s \bar q \sigma G q \rangle \over 4608 \pi^4 }
+ { 5 \langle g_s^2 G G \rangle \langle \bar q q \rangle \langle \bar s s \rangle \over 864 \pi^2 }
+ { m_s^4 \langle \bar q q \rangle^2 \over 12 \pi^2 }
\\ \nonumber && - { 7 m_s^2 \langle \bar q q \rangle \langle g_s \bar q \sigma G q \rangle \over 48 \pi^2 }
- { m_s^2 \langle \bar q q \rangle \langle g_s \bar s \sigma G s \rangle \over 48 \pi^2 }
+ { 5 m_s \langle \bar q q \rangle^2 \langle \bar s s \rangle \over 9 }
+ { \langle g_s \bar q \sigma G q \rangle \langle g_s \bar s \sigma G s \rangle \over 96 \pi^2 }
\Big ),
\end{eqnarray}
\begin{eqnarray}\nonumber
\Pi^{M(q s \bar q \bar s)}_{4\mu}(M_B^2) &=& \int^{s_0}_{4m_s^2} \Bigg [
{ 1 \over 12288 \pi^6 } s^4
- { 3 m_s^2 \over 1024 \pi^6 } s^3
+ \Big (
{ \langle g_s^2 G G \rangle \over 18432 \pi^6 }
+ { 9 m_s^4 \over 512 \pi^6 } 
+ { m_s \langle \bar s s \rangle \over 64 \pi^4 }
\Big ) s^2 
\\ \nonumber && + \Big (
{ m_s^2 \langle g_s^2 G G \rangle \over 2048 \pi^6 }
- { m_s^3 \langle \bar s s \rangle \over 8 \pi^4 }
+ { m_s \langle g_s \bar q \sigma G q \rangle \over 192 \pi^4 }
- { m_s \langle g_s \bar s \sigma G s \rangle \over 64 \pi^4 }
+ { \langle \bar q q \rangle^2 \over 24 \pi^2 }
+ { \langle \bar s s \rangle^2 \over 24 \pi^2 }
\Big ) s
\\ \nonumber && + \Big (
{ m_s \langle g_s^2 G G \rangle \langle \bar q q \rangle \over 128 \pi^4 }
- { m_s \langle g_s^2 G G \rangle \langle \bar s s \rangle \over 256 \pi^4 }
- { 3 m_s^2 \langle \bar q q \rangle^2 \over 4 \pi^2 }
+ { m_s^2 \langle \bar s s \rangle^2 \over 16 \pi^2 }
+ { \langle \bar q q \rangle \langle g_s \bar q \sigma G q \rangle \over 16 \pi^2 }
- { \langle \bar q q \rangle \langle g_s \bar s \sigma G s \rangle \over 48 \pi^2 }
\\ \nonumber && - { \langle \bar s s \rangle \langle g_s \bar q \sigma G q \rangle \over 48 \pi^2 }
+ { \langle \bar s s \rangle \langle g_s \bar s \sigma G s \rangle \over 16 \pi^2 }
\Big )
\Bigg ] e^{-s/M_B^2} ds
+ \Big (
{ m_s^3 \langle g_s^2 G G \rangle \langle \bar s s \rangle \over 1152 \pi^2 }
+ { 5 m_s \langle g_s^2 G G \rangle \langle g_s \bar q \sigma G q \rangle \over 2304 \pi^4 }
\\ \nonumber && - { 5 m_s \langle g_s^2 G G \rangle \langle g_s \bar s \sigma G s \rangle \over 4608 \pi^4 }
+ { 5 \langle g_s^2 G G \rangle \langle \bar q q \rangle^2 \over 1728 \pi^2 }
- { 5 \langle g_s^2 G G \rangle \langle \bar q q \rangle \langle \bar s s \rangle \over 432 \pi^2 }
+ { 5 \langle g_s^2 G G \rangle \langle \bar s s \rangle^2 \over 1728 \pi^2 }
+ { m_s^4 \langle \bar q q \rangle^2 \over 4 \pi^2 }
\\ \nonumber && - { m_s^2 \langle \bar q q \rangle \langle g_s \bar q \sigma G q \rangle \over 2 \pi^2 }
+ { m_s^2 \langle \bar q q \rangle \langle g_s \bar s \sigma G s \rangle \over 24 \pi^2 }
+ { 5 m_s \langle \bar q q \rangle^2 \langle \bar s s \rangle \over 3 }
+ { \langle g_s \bar q \sigma G q \rangle^2 \over 64 \pi^2 }
- { \langle g_s \bar q \sigma G q \rangle \langle g_s \bar s \sigma G s \rangle \over 48 \pi^2 }
\\ \nonumber && + { \langle g_s \bar s \sigma G s \rangle^2 \over 64 \pi^2 }
\Big ),
\end{eqnarray}

\end{widetext}

%

\end{document}